\documentclass[10pt,journal,compsoc]{IEEEtran}

\usepackage{cite}
\usepackage{amsmath,amssymb,amsfonts}
\usepackage{algorithm}
\usepackage{algpseudocode}
\usepackage{graphicx}
\usepackage{xcolor}
\usepackage{textcomp}
\usepackage{array}
\usepackage{soul}
\usepackage{subfig}
\usepackage{multirow}
\usepackage[nolist]{acronym}
\usepackage{enumitem}
\usepackage{siunitx}
\usepackage{caption}
\usepackage{pgfplots}
\usepackage{tikz}
\usetikzlibrary{patterns}
\usepackage{pdfpages}
\usepackage{booktabs}
\usepackage{adjustbox}
\usepackage{tabularx}
\usepackage{ragged2e}
\usepackage{multicol}
\usepackage{pifont}
\usepackage{ragged2e} % for \RaggedRight
\newcolumntype{L}{>{\RaggedRight\arraybackslash}X}

\pgfplotsset{compat=1.17}

\def\BibTeX{{\rm B\kern-.05em{\sc i\kern-.025em b}\kern-.08em
    T\kern-.1667em\lower.7ex\hbox{E}\kern-.125emX}}
\AtBeginDocument{\definecolor{tmlcncolor}{cmyk}{0.93,0.59,0.15,0.02}\definecolor{NavyBlue}{RGB}{0,86,125}}

 \usepackage{bm}

\newcommand{\figref}[1]{\figurename~{\ref{#1}}}

\newcommand{\tabref}[1]{Table~\ref{#1}}

\begin{acronym}
    \acro{GAI}{Generative Artificial Intelligence}
    \acro{xApp}{eXtended Application}
    \acro{QoS}{Quality of Service}
    \acro{5G}{Fifth Generation}
    \acro{6G}{Sixth Generation}
    \acro{MIMO}{Multiple-Input Multiple-Output}
    \acro{ML}{Machine Learning}
    \acro{AI}{Artificial Intelligence}
    \acro{CSI}{Channel State Information}
    \acro{MAC}{Medium Access Control}
    \acro{O-RAN}{Open Radio Access Network}
    \acro{RAN}{Radio Access Network}
    \acro{C-RAN}{Cloud Radio Access Network}
    \acro{vRAN}{Virtualized Radio Access Network}
    \acro{PDCP}{Packet Data Convergence Protocol}
    \acro{SDAP}{Service Data Adaptation Protocol}
    \acro{eMBB}{Enhanced Mobile Broadband}
    \acro{URLLC}{Ultra-Reliable Low Latency Communications}
    \acro{RU}{Radio Unit}
    \acro{DU}{Distributed Unit}
    \acro{CU}{Central Unit}
    \acro{CP}{Control Plane}
    \acro{UP}{User Plane}
    \acro{MANO}{Management and Orchestration}
    \acro{NFV}{Network Functions Virtualization}
    \acro{RIC}{Radio Access Network Intelligent Controller}
    \acro{near-RT}{near-real-time}
    \acro{non-RT}{non-real-time}
    \acro{DL}{Deep Learning}
    \acro{RL}{Reinforcement Learning}
    \acro{DRL}{Deep Reinforcement Learning}
    \acro{VAE}{Variational Autoencoder}
    \acro{UE}{User Equipment}
    \acro{PRB}{Physical Resource Block}
    \acro{SINR}{Signal to Interference \& Noise Ratio}
    \acro{ESA}{Exhaustive Search Algorithm}
    \acro{DNN}{Deep Neural Network}
    \acro{MDP}{Markov Decision Process}
    \acro{DQN}{Deep Q-Network}
    \acro{BS}{Base Station}
    \acro{KL}{Kullback-Leibler}
    \acro{MSE}{Mean Squared Error}
    \acro{MAE}{Mean Absolute Error}
    \acro{mMTC}{Massive Machine-Type Communications}
    \acro{FL}{Federated Learning}
    \acro{TL}{Transfer Learning}
    \acro{NLOS}{non-line-of-sight}
    \acro{RB}{Resource Block}
    \acro{AMF}{Access and Mobility Management Function}
    \acro{UPF}{User Plane Function}
    \acro{OPEX}{Operational Expenditures}
    \acro{LLM}{Large Language Model}
    \acro{MARL}{Multi-Agent Reinforcement Learning}
    \acro{FDRL}{Federated Deep Reinforcement Learning}
    \acro{GAN}{Generative Adversarial Network}
    \acro{DM}{Diffusion Model}
\end{acronym}

\begin{document}

\title{Generative AI for O-RAN Slicing: A Semi-Supervised Approach with VAE and Contrastive Learning}

\author{Salar Nouri\;\;\;
	Mojdeh Karbalaee Motalleb \;\;\;
	Vahid Shah-Mansouri \; \; \; 
	Seyed Pooya Shariatpanahi\\
	 School of Electrical and Computer Engineering, College of Engineering, University of Tehran, Tehran, Iran\\
	\texttt{\{salar.nouri, mojdeh.karbalaee, vmansouri, p.shariatpanahi\}@ut.ac.ir}}

% \markboth{Journal of \LaTeX\ Class Files,~Vol.~14, No.~8, August~2015}%
% {Shell \MakeLowercase{\textit{et al.}}: Bare Demo of IEEEtran.cls for Computer Society Journals}


% ********************** Apply Different Parts ******************** %
\IEEEtitleabstractindextext{
\begin{abstract}
This paper introduces a novel generative AI (GAI)-driven, unified semi-supervised learning architecture for optimizing resource allocation and network slicing in O-RAN. Termed Generative Semi-Supervised VAE-Contrastive Learning, our approach maximizes the weighted user equipment (UE) throughput and allocates physical resource blocks (PRBs) to enhance the quality of service for eMBB and URLLC services. The GAI framework utilizes a dedicated xApp for intelligent power control and PRB allocation. This integrated GAI model synergistically combines the generative power of a VAE with contrastive learning to achieve robustness in an end-to-end trainable system. It is a semi-supervised training approach that concurrently optimizes supervised regression of resource allocation decisions (i.e., power, UE association, PRB) and unsupervised contrastive objectives. This intrinsic fusion improves the precision of resource management and model generalization in dynamic mobile networks. We evaluated our GAI methodology against exhaustive search and deep Q-Network algorithms using key performance metrics. Results show our integrated GAI approach offers superior efficiency and effectiveness in various scenarios, presenting a compelling GAI-based solution for critical network slicing and resource management challenges in next-generation O-RAN systems.
\end{abstract}

\begin{IEEEkeywords}
Generative AI, Variational Autoencoder, Contrastive Learning, \ac{O-RAN} Slicing, Resource Allocation
\end{IEEEkeywords}

}

\maketitle
\IEEEdisplaynontitleabstractindextext

\IEEEpeerreviewmaketitle
\IEEEraisesectionheading{\section{Introduction}\label{sec:introduction}}
% \section{INTRODUCTION}
% \label{introduction}

\IEEEPARstart{W}ireless communication has evolved from a convenience to a necessity, driven by the increasing demand for faster and more reliable connectivity across diverse \ac{QoS} requirements \cite{larsen2018surveysplit}. The advent of \ac{5G} marks a significant leap, enabling applications such as augmented reality and autonomous vehicles \cite{nassef2022survey}, while laying the foundation for \ac{6G}, which aims to further redefine global connectivity.

To meet the growing complexity of future networks, 3GPP begins integrating \ac{AI} into core \ac{RAN} functions such as beam management, positioning, and \ac{CSI} feedback, initiating a shift towards intelligent and adaptive systems. As technologies like extreme \ac{MIMO}, deep receivers, and deep schedulers emerge, \ac{AI} becomes essential to handle real-time data-driven tasks such as \acp{PRB} allocation and power control in both uplink and downlink \cite{zhang2019overview, zhao2023multi, lin2023overview}.

Network slicing and \ac{O-RAN} are critical enablers in both \ac{5G} and \ac{6G}, offering the flexibility to support services like \ac{eMBB}, \ac{URLLC}, and \ac{mMTC}, each with distinct performance demands \cite{popovski20185g, zhao2023multi}. In \ac{6G}, network slicing will extend to advanced use cases such as holographic communication and tactile Internet, ensuring scalable and adaptable resource management \cite{huang2023opportunistic, al2023resource}. At the same time, \ac{O-RAN}—with its open interfaces, disaggregation of hardware and software—provides a modular, cost-effective platform for embedding \ac{AI}-based control and enabling multi-vendor interoperability \cite{abdalla2022toward}.

These architectures are foundational for realizing sophisticated \ac{AI}-driven resource allocation strategies. As network demands grow and \ac{QoS} requirements diversify, highly adaptive intelligent solutions for dynamic scheduling, power control, and \ac{PRB} management become increasingly paramount. Effectively integrating such advanced \ac{AI} within flexible frameworks such as network slicing and \ac{O-RAN} is crucial to ensure future wireless systems are not only efficient and responsive but also proactively service-aware, making sophisticated \ac{AI} capabilities indispensable for unlocking the full potential of \ac{6G} \cite{shehzad2022artificial, motalleb2023moving}.

The inherent complexity and dynamism of these emerging wireless environments have spurred the adoption of \ac{AI} algorithms, as conventional optimization methods often lack the agility to cope with rapidly evolving network states and diverse performance criteria. Although learning-based paradigms, such as \ac{DL}, have demonstrated potential for specific \ac{RAN} optimization tasks, their application to multifaceted resource allocation in \ac{O-RAN} slicing often encounters significant hurdles. These typically include the need for extensive, task-specific agent engineering, computationally intensive training regimens that usually start from scratch, and practical limitations related to sample inefficiency and suboptimal generalization to novel network conditions or previously unseen service demands. Such drawbacks critically hinder the development of truly autonomous systems capable of generating robust, adaptive, and near-optimal resource management strategies in real time.

The previously discussed limitations of conventional \ac{AI} methods in achieving adaptable and data-efficient resource allocation for dynamic networks underscore the need for novel \ac{GAI} paradigms. To this end, we propose an innovative \ac{GAI} framework, termed Generative Semi-Supervised \ac{VAE}-Contrastive Learning (SS-VAE), designed explicitly for intelligent \ac{O-RAN} network slicing. This framework is realized through a unified architecture that intrinsically integrates the distributional learning and generative capabilities of a \ac{VAE} \cite{kingma2013auto} with the potent representation learning power of contrastive learning principles \cite{chen2020simple}. Trained end-to-end using a semi-supervised strategy, our SS-VAE model is designed to learn complex resource distributions and generate optimized allocation decisions with superior generalization and data efficiency.

The synergistic design of our Generative SS-VAE framework offers distinct advantages for \ac{GAI}-driven resource optimization in \ac{O-RAN}. The \ac{VAE} core excels at modeling the underlying probability distributions of network states and resource utilization, which is crucial for managing the inherent uncertainty and can be leveraged to handle incomplete data by inferring missing values from learned patterns. Its generative capability, central to our \ac{GAI} approach, allows the model to proactively construct or refine resource allocation strategies, finely tuned to diverse \ac{QoS} demands within \ac{O-RAN} slices. This is powerfully augmented by the integrated contrastive learning mechanism, which compels the model to learn highly informative and robust latent representations from both limited labeled data and abundant unlabeled data. Such a focus on representation quality, integral to our semi-supervised methodology, significantly enhances sample efficiency and fortifies the model's ability to generalize to unseen network conditions and resist noise or outliers, making it exceptionally suited for the dynamic \ac{O-RAN} environment.

Our proposed SS-VAE algorithm is designed for practical implementation within the \ac{O-RAN} architecture. It is envisioned for deployment as an \ac{xApp} operating on the near-real-time \ac{RIC}. This strategic placement enables the SS-VAE \ac{xApp} to dynamically adjust resource allocation by leveraging real-time network telemetry and user demand information, thereby facilitating efficient, adaptive, and \ac{QoS}-aware management of network slices.

\subsection{Contribution}
This paper makes significant contributions to the field of intelligent resource allocation and network slicing within the \ac{O-RAN} architecture, explicitly targeting the distinct demands of \ac{eMBB} and \ac{URLLC} services. We introduce and rigorously evaluate a novel \ac{GAI} driven solution. The primary contributions of this work are as follows.

\begin{itemize}
    \item We present a detailed system model and formulate the resource allocation problem in \ac{O-RAN} environments, aiming to optimize weighted throughput via efficient assignment of transmission power and \acp{PRB} in the downlink. This formulation explicitly considers the unique \ac{QoS} requirements and service priorities of \ac{eMBB} and \ac{URLLC} services. It is designed for implementation within an \ac{xApp}-based architecture to enhance the flexibility of \ac{AI}-driven network management.

    \item We propose, design and implement a novel \ac{GAI} framework to optimize resource allocation in \ac{O-RAN}. This framework employs a unified deep learning architecture that combines a \ac{VAE}—which captures complex data patterns and can generate resource allocation solutions—with contrastive learning to learn robust and meaningful feature representations. The entire model is trained end-to-end using a semi-supervised approach, which improves both data efficiency and overall performance. This enables effective generalization across various O-RAN environment scenarios. This makes the framework highly adaptable and practical for real-world, dynamic O-RAN systems.

    \item We demonstrate through extensive simulations and comparative analysis against multiple benchmarks—including the exhaustive search algorithm (ESA) and deep reinforcement learning—that our proposed \ac{AI}-driven approach achieves near-optimal resource allocation performance with substantially lower computational complexity than state-of-the-art methods. The results validate the robustness, computational efficiency, and scalability of our \ac{AI} solution, affirming its suitability for real-time resource management in dynamic \ac{O-RAN} systems. Furthermore, the principles of our semi-supervised generative model pave the way for its application in future multi-vendor \ac{O-RAN} architectures and evolving service paradigms.
\end{itemize}

\subsection{Structure of the Paper}

The paper is structured as follows: Section \ref{Background} provides a brief overview of the \ac{O-RAN} architecture and its components relevant to intelligent control. Section \ref{literature_review} reviews related works in resource allocation, particularly focusing on \ac{AI}-based approaches in \ac{O-RAN}. In Section \ref{system_model}, we introduce the system model for the resource allocation challenge in the \ac{O-RAN} architecture and formally define the problem as an optimization task. Section \ref{Proposed_scheme} elaborates on our novel Generative AI-driven resource allocation framework, detailing its underlying model architecture and training methodology. Section \ref{simulation_results} presents the numerical results from our performance evaluation, comparing our approach against benchmark algorithms. Section \ref{complexity} provides an analysis of the computational complexity of the employed algorithms. Finally, Section \ref{conclusions} offers concluding remarks and outlines potential future research directions. 
For clarity, the following section lists the main acronyms used throughout the paper.

\section*{List of Acronyms and Abbreviations}
\begin{description}[style=unboxed, leftmargin=2.5cm, labelsep=1em]
    \item[\textbf{5G}] Fifth Generation
    \item[\textbf{6G}] Sixth Generation
    \item[\textbf{AI}] Artificial Intelligence
    \item[\textbf{AMF}] Access and Mobility Management Function
    \item[\textbf{BS}] Base Station
    \item[\textbf{C-RAN}] Cloud Radio Access Network
    \item[\textbf{CP}] Control Plane
    \item[\textbf{CSI}] Channel State Information
    \item[\textbf{CU}] Central Unit
    \item[\textbf{DL}] Deep Learning
    \item[\textbf{DM}] Diffusion Model
    \item[\textbf{DNN}] Deep Neural Network
    \item[\textbf{DQN}] Deep Q-Network
    \item[\textbf{DRL}] Deep Reinforcement Learning
    \item[\textbf{DU}] Distributed Unit
    \item[\textbf{eMBB}] Enhanced Mobile Broadband
    \item[\textbf{ESA}] Exhaustive Search Algorithm
    \item[\textbf{FDRL}] Federated Deep Reinforcement Learning
    \item[\textbf{FL}] Federated Learning
    \item[\textbf{GAI}] Generative Artificial Intelligence
    \item[\textbf{GAN}] Generative Adversarial Network
    \item[\textbf{KL}] Kullback-Leibler
    \item[\textbf{LLM}] Large Language Model
    \item[\textbf{MAC}] Medium Access Control
    \item[\textbf{MAE}] Mean Absolute Error
    \item[\textbf{MANO}] Management and Orchestration
    \item[\textbf{MARL}] Multi-Agent Reinforcement Learning
    \item[\textbf{MDP}] Markov Decision Process
    \item[\textbf{MIMO}] Multiple-Input Multiple-Output
    \item[\textbf{ML}] Machine Learning
    \item[\textbf{mMTC}] Massive Machine-Type Communications
    \item[\textbf{MSE}] Mean Squared Error
    \item[\textbf{near-RT}] Near-Real-Time
    \item[\textbf{NFV}] Network Functions Virtualization
    \item[\textbf{NLOS}] Non-Line-of-Sight
    \item[\textbf{non-RT}] Non-Real-Time
    \item[\textbf{O-RAN}] Open Radio Access Network
    \item[\textbf{OPEX}] Operational Expenditures
    \item[\textbf{PDCP}] Packet Data Convergence Protocol
    \item[\textbf{PRB}] Physical Resource Block
    \item[\textbf{QoS}] Quality of Service
    \item[\textbf{RAN}] Radio Access Network
    \item[\textbf{RB}] Resource Block
    \item[\textbf{RIC}] Radio Access Network Intelligent Controller
    \item[\textbf{RL}] Reinforcement Learning
    \item[\textbf{RU}] Radio Unit
    \item[\textbf{SDAP}] Service Data Adaptation Protocol
    \item[\textbf{SINR}] Signal to Interference \& Noise Ratio
    \item[\textbf{TL}] Transfer Learning
    \item[\textbf{UE}] User Equipment
    \item[\textbf{UP}] User Plane
    \item[\textbf{UPF}] User Plane Function
    \item[\textbf{URLLC}] Ultra-Reliable Low Latency Communications
    \item[\textbf{VAE}] Variational Autoencoder
    \item[\textbf{vRAN}] Virtualized Radio Access Network
    \item[\textbf{xApp}] eXtended Application
\end{description}
%%%%%%%%%%%%%%%%%%%%%%%%%%%%%%%%%%%%%%%%%%%%%%%%%%%%%%%%%%%%%%%%%%%%%%%%%%%%%%%%%%%%%%%%%%%%%%%%

\section{Background}
\label{Background}
\begin{figure}[t]
	\centerline{\includegraphics[width=0.5\textwidth]{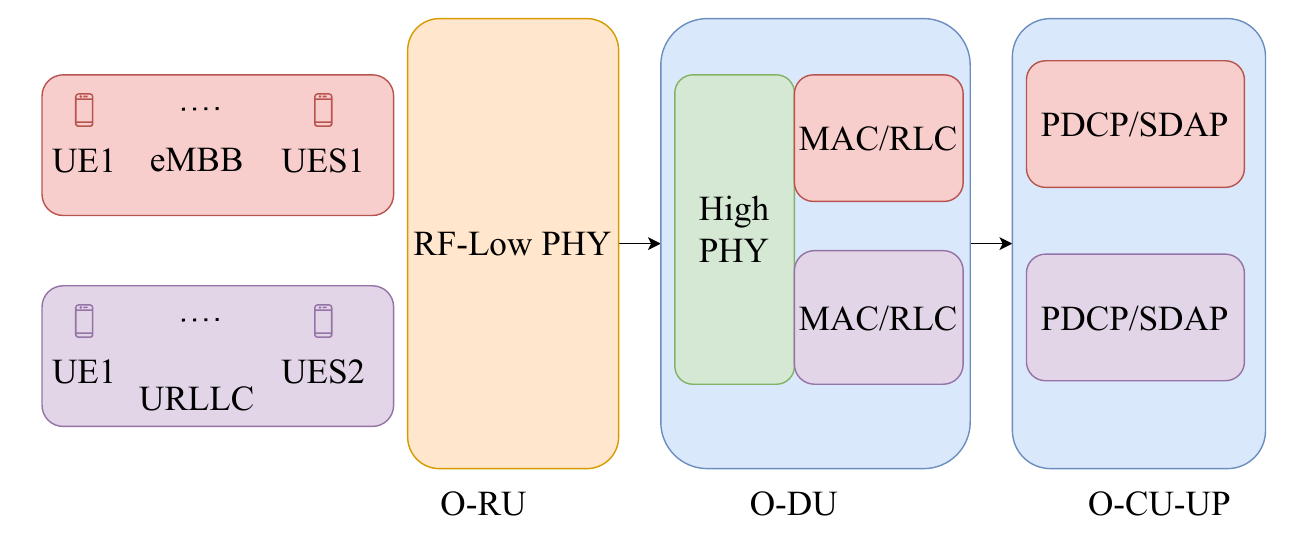}}
	\caption{\ac{O-RAN} network slicing}
	\label{fig:slicing}
\end{figure}

The \ac{O-RAN} Alliance has created an innovative \ac{RAN} framework designed to enable an open, intelligent, virtualized, and interoperable \ac{RAN}, which is crucial for affordable, next-generation wireless networks \cite{oranwhitepaper}.
The \ac{O-RAN} architecture combines \ac{C-RAN} and \ac{vRAN} to create open, interoperable, and flexible mobile networks that support both \ac{5G} and future \ac{6G} requirements. The core components of \ac{O-RAN} include the Open Radio Unit (O-RU), the Open Distributed Unit (O-DU), and the Open Centralized Unit (O-CU), each responsible for specific layers of the protocol stack. The O-RU handles radio frequency (RF) and lower physical (PHY) functions. The O-DU processes the medium access control (MAC), radio link control (RLC), and higher PHY layers. Meanwhile, the O-CU manages the \ac{PDCP} and \ac{SDAP} and is further divided into the \ac{CP} and \ac{UP} for efficient signaling and traffic routing \cite{Nokiawhitepaper, oranwhitepaper}.
Beyond these disaggregated data plane components, key \ac{O-RAN} architectural pillars include Orchestration and Automation, the O-Cloud, and notably, the \ac{RAN} \ac{RIC}. The \ac{RIC} is pivotal in enabling intelligent network operations, facilitating near-real-time and non-real-time optimization and learning-based control over the \ac{RAN}.

In the subsequent section, we provide a brief overview of the key methods and characteristics employed in the \ac{O-RAN} system, which improve its adaptability and efficiency.

\subsection{Network Slicing in O-RAN}
A cornerstone of \ac{5G}/\ac{6G} models, network slicing facilitates the creation of customized virtual networks on demand that operate on shared infrastructure. This end-to-end capability is achieved by logically separating and managing resources across three key domains: the \ac{RAN}, the core, and the transport network. In the RAN domain, slicing involves partitioning radio resources, such as \acp{PRB}, and instantiating dedicated virtual network functions for different services, including MAC/RLC in the O-DU and \ac{PDCP}/\ac{SDAP} in the O-CU, as illustrated in \figref{fig:slicing}. In the core network, key nodes, such as the \ac{UPF} and \ac{AMF}, are virtualized and isolated to create slices that accommodate specific service-level agreements. The transport network is sliced to create dedicated data pathways, ensuring that the performance and \ac{QoS} for these diverse connections are guaranteed. The \ac{O-RAN} framework, with its inherent intelligence and virtualization, serves as a crucial enabler for sophisticated \ac{RAN} slicing, which is indispensable to perform comprehensive end-to-end network slicing \cite{javadpour2023reinforcement, motalleb2022resource}.

\subsection{Radio Intelligent Controller (\ac{RIC})}
The \ac{RIC} is functionally divided into two layers: the \ac{near-RT} and the \ac{non-RT} \ac{RIC}. The \ac{near-RT} \ac{RIC}, designed for control loops operating within 10 milliseconds to 1 second latency, hosts modular xApps that enable fast and responsive control over \ac{RAN} functions. These \acp{xApp} handle tasks such as interference mitigation, handover control, and dynamic resource allocation, interacting with the \ac{RAN} in real time. They are fundamental to enabling service-level customization, particularly through the use of network slicing.
In contrast, the \ac{non-RT} \ac{RIC}, typically situated within the broader Orchestration and Automation framework, manages operations with latency requirements exceeding 1 second. It supports functions like policy management, network analytics, and \ac{AI} model training by hosting rApps. These applications analyze historical data, orchestrate the training of \ac{AI} models, and generate strategic policies or guidance for the \ac{near-RT} \ac{RIC} and its \acp{xApp} \cite{bonati2021intelligence, motalleb2025towards}.

Together, the \ac{non-RT} \ac{RIC} and \ac{near-RT} \ac{RIC} establish a hierarchical intelligent control loop. Typically, \ac{AI} models are trained or designed in the \ac{non-RT} \ac{RIC} using historical data and network policies and then deployed or used by \acp{xApp} in the \ac{near-RT} \ac{RIC} for inference and real-time action. This synergy enables predictive and adaptive \ac{RAN} behavior, significantly enhancing automation in dynamic network environments.
Both \acp{xApp} and rApps are designed to be modular and leverage open interfaces, which encourages third-party innovation and accelerates the deployment of \ac{AI}-driven functionalities, thereby promoting multi-vendor interoperability within the \ac{O-RAN} ecosystem.

\subsection{Artificial Intelligence (\ac{AI}) in O-RAN}
The integration of \ac{AI} is a foundational tenet of the \ac{O-RAN} architecture, designed to introduce intelligence and autonomy into network operations. This approach is crucial for enhancing \ac{QoS} through dynamic optimization and reducing \ac{OPEX} through automated management. The application of \ac{AI} is envisioned across a spectrum of \ac{RAN} use cases, from real-time resource allocation and traffic steering to anomaly detection and cybersecurity.

To address these challenges, the \ac{O-RAN} framework is equipped to leverage a diverse range of \ac{AI} methodologies. Supervised learning enables predictive capabilities, such as traffic forecasting, by training on labeled historical data. In contrast, unsupervised learning identifies latent patterns in unlabeled data for tasks such as user clustering or threat detection. For autonomous control, \ac{RL} and its deep-learning variant (\ac{DRL}) allow agents to derive optimal policies through direct interaction with the network environment. To address privacy concerns in multi-agent environments, \ac{FL} facilitates collaborative model training on distributed data without centralizing sensitive information. Further advancing these capabilities, \ac{GAI} models like \acp{VAE} can learn data distributions to generate synthetic training data or even propose new resource management solutions directly. These core techniques can be augmented by \acp{LLM}, which provides a semantic understanding of unstructured data to enhance the overall intelligence of the system's decision-making.

These \ac{AI} functionalities are operationally hosted within the \ac{O-RAN} Intelligent Controllers (\ac{RIC}). The \ac{non-RT} \ac{RIC} typically manages functions with relaxed latency constraints, such as offline model training or acting as the central server in a Federated Learning setup. Conversely, the \ac{near-RT} \ac{RIC} executes tasks requiring rapid response, including model inference for sub-second control loops and the distributed training aspects of \ac{RL} and \ac{FL}. This architectural division allows \ac{O-RAN} to effectively support the entire \ac{AI} lifecycle, from model creation in the \ac{non-RT} \ac{RIC} to real-time deployment and execution in the \ac{near-RT} \ac{RIC}.
\section{Related Works}
\label{literature_review}

As highlighted in Section \ref{sec:introduction}, ensuring reliable connectivity for devices with diverse \ac{QoS} requirements presents significant challenges for cellular operators, especially in the complex and demanding landscape of \ac{5G} and emerging \ac{6G} networks \cite{feng2020dynamic}. As a result, network slicing has become a key focus of research in cellular networks. With the increasing complexity and diversity of services, methodologies for resource allocation have undergone significant evolution. They have shifted from traditional mathematical optimization and heuristics to more adaptive and intelligent AI-driven approaches \cite{feng2020dynamic, motalleb2019joint, kazemifard2021minimum, hurtado2022deep, adhikari2023resource}.

\subsection{Conventional Optimization and Heuristic Approaches}
Resource allocation in wireless networks has been addressed through formal mathematical optimization and heuristic algorithms \cite{feng2020dynamic, motalleb2019joint, kazemifard2021minimum, chen2023adaptive}. Algorithms such as Linear and Non-Linear Programming provide a rigorous framework for defining these problems, and for specific cases like convex problems, can even yield globally optimal solutions efficiently. When exact solutions are computationally infeasible due to the NP-hard nature of many practical scenarios involving discrete decisions, a wide range of heuristic and meta-heuristic algorithms, including greedy methods, simulated annealing, and genetic algorithms, have been employed. These methods aim to find good, albeit often suboptimal, solutions quickly, providing a practical alternative to complex optimization.

Despite their utility, these conventional approaches have significant limitations in the context of dynamic, sliced \ac{O-RAN} environments. Their primary failing is the computational complexity that makes them too slow for the near-real-time control loops required by the \ac{O-RAN} \ac{RIC}. Furthermore, these methods typically rely on static models and lack the agility to adapt to the highly dynamic nature of wireless channels, traffic loads, and user mobility. Their performance is tied to the accuracy of underlying network models, and heuristics can be brittle, performing poorly when conditions deviate from their design assumptions. Crucially, they do not learn from operational data to improve future decisions, which makes them ill-suited for the data-rich, learning-driven paradigm enabled by \ac{O-RAN}'s open interfaces.

\subsection{Discriminative AI and Reinforcement Learning Paradigms}
The limitations of conventional optimization have driven research into \ac{AI}-driven solutions, with \ac{DRL} emerging as a significant approach for dynamic resource allocation. \ac{DRL} combines the function approximation capabilities of \ac{DNN} with the trial-and-error framework of \ac{RL}, enabling agents to learn optimal policies through interaction with their environment. Various \ac{DRL} algorithms, including value-based methods like \ac{DQN} and actor-critic methods such as Soft Actor-Critic, have been utilized in wireless resource management \cite{lotfi2llm}. In complex systems like \ac{O-RAN}, \ac{MARL} allows multiple agents to learn simultaneously, optimizing local and global objectives. \ac{FL} has also gained importance as a distributed machine learning paradigm that facilitates collaborative training without centralizing data, thus preserving privacy. The combination of \ac{DRL} and \ac{FL}, termed \ac{FDRL}, allows distributed agents to enhance \ac{DRL} model building. They offer advantages such as adaptability, reduced reliance on precise network models, the ability to manage high-dimensional state spaces, and enhanced data privacy \cite{salama2025fedora}. However, \ac{DRL} faces challenges such as sample inefficiency, long training times, and poor generalization in new scenarios. Additionally, designing practical reward functions is complex, and the "black box" nature of deep networks raises interpretability concerns. Moreover, \ac{FL} struggles with non-identically distributed data and communication overhead. \ac{DRL} effectively learns reactive policies but does not model the generative processes or distributional characteristics of optimal solutions, highlighting the need to explore generative AI approaches to understand these underlying data distributions better.

\subsection{The Rise of Generative AI in Wireless Network Optimization}
A new frontier in \ac{AI}-driven network optimization is emerging with the application of \ac{GAI}. Unlike discriminative models that learn decision boundaries or \ac{RL} agents that learn reactive policies, \ac{GAI} models are designed to learn the underlying probability distribution of data. This allows them to generate new, synthetic data instances or solutions that exhibit similar characteristics to the training data \cite{zhao2024generative}. The unique capabilities of \ac{GAI} offer several advantages for complex optimization problems, such as \ac{O-RAN} resource allocation, including the ability to proactively generate novel and optimized solutions, augment limited training datasets, and robustly model the inherent uncertainty of wireless environments by capturing complex data distributions \cite{uslu2025generative}.

This generative capability is realized through several prominent model architectures. \acp{VAE} are adept at learning compressed, meaningful representations of high-dimensional network data, making them suitable for both generating solutions and performing inference [1, 49]. \acp{GAN} uses a competitive training process and has been primarily explored for data augmentation, creating realistic synthetic channel data or traffic patterns to enhance the robustness of other AI models [35, 45]. More recently, \acp{DM} have shown promise for generating high-fidelity samples and are being investigated for imitating expert allocation policies \cite{chen2024big}. The choice of model often depends on the specific task, whether it is direct solution generation, data augmentation, or learning latent representations for other processes.

\subsection{Identifying the Research Gap: Positioning the Proposed SS-VAE Framework}

Despite notable advancements, current resource allocation methods in sliced \ac{O-RAN} environments still face significant challenges. Conventional optimization and heuristic approaches often lack the speed required for real-time control and struggle to adapt to dynamic network conditions. \ac{DRL} provides adaptability but suffers from poor sample efficiency and limited generalization, as it typically learns reactive policies without fully modeling the optimal solution distribution. Advanced techniques, such as \ac{FL} and \ac{TL}, offer some benefits but do not adequately address the need to learn from limited labeled data while exploring the vast solution space of NP-hard allocation problems.

There is a pressing need for approaches that can learn robust representations from scarce or mixed-quality data, model complex uncertainties, and explore the solution space in a generative manner to find near-optimal resource allocation strategies. Specifically, this is essential for coordinating multiple \acp{xApp} within dynamic \ac{O-RAN} environments. The challenge is heightened by the limited generalizability of existing models across diverse deployment conditions, including varying numbers of \ac{UE}, fluctuating traffic patterns, and heterogeneous network configurations. This limits their scalability and effectiveness in real-world scenarios.

To address these issues, we propose the Generative Semi-Supervised \ac{VAE}-Contrastive Learning (SS-VAE) framework, which integrates key \ac{AI} concepts to fill this gap. At its core, the \ac{VAE} learns the underlying probability distributions of network states and optimal resource allocations, enabling proactive generation and refinement of allocation decisions. This generative capability is vital for navigating the complex solution landscape of joint resource allocation, moving beyond the reactive tendencies of \ac{DRL} and the data demands of purely supervised learning.
\tabref{tab:comparative_analysis} summarizes the key characteristics, contributions, and limitations of selected related works alongside the proposed SS-VAE approach.

\begin{table*}[!t]
\centering
\caption{Comparative Analysis of AI-Driven Resource Allocation Methods in O-RAN}
\label{tab:comparative_analysis}
\newcolumntype{L}{>{\RaggedRight\arraybackslash}X} % For left-aligned text wrapping
\begin{tabularx}{\textwidth}{@{}l L L L L L@{}}
\toprule
\textbf{Reference} & \textbf{Problem Domain} & \textbf{Methodology} & \textbf{Key Contribution(s)} & \textbf{Limitation(s)} & \textbf{Primary Differentiator vs. This Work} \\
\midrule

\textbf{SS-VAE} &
Multi-resource (power, \ac{PRB}, \ac{UE} assoc.) allocation for \ac{eMBB}/\ac{URLLC} slicing in \ac{O-RAN}. &
Semi-Supervised Generative \ac{AI} (VAE with Contrastive Learning). &
\begin{itemize}[leftmargin=*,noitemsep] \item Novel unified GAI framework. \item Data-efficient via semi-supervision. \item Generative exploration of solution space. \end{itemize} &
Requires initial limited labels from a high-complexity solver. &
\textit{(Baseline for comparison)} \\
\hline

Li et al. \cite{li2019deep} &
Proactive resource allocation considering UE mobility in HetNets. &
DRL. &
Adapts proactively to user mobility. &
Standard DRL challenges (e.g., sample complexity, generalization). &
Learns a reactive policy, whereas our work is generative, learning the solution distribution. \\
\hline

Wu et al. \cite{wu2020dynamic} &
Joint radio and computation resource allocation for vehicular RAN slicing. &
DRL. &
Dynamically allocates both radio and compute resources. &
Relies on reactive DRL policy; generalization across diverse slice types. &
Reactive DRL policy vs. our generative exploration of the joint resource space. \\
\hline

Wang et al. \cite{wang2022self} &
Resource management for O-RAN RU and DU entities. &
Self-Play DRL. &
Achieves adaptive decision-making through a self-play paradigm. &
Scalability and generalization of self-play in a complex O-RAN environment. &
Game-theoretic DRL policy vs. our GAI approach to learning underlying data characteristics. \\
\hline

Ndikumana et al. \cite{ndikumana2023federated} &
Joint task offloading and fronthaul routing in O-RAN to reduce delay. &
FL + DRL. &
Ensures data privacy via FL while optimizing routing decisions. &
High complexity in coordinating FL and DRL; potential convergence issues with non-IID data. &
Augments a reactive DRL policy with FL, while our method is a unified generative model. \\
\hline

Mhatre et al. \cite{mhatre2025transfer} &
Efficient resource management in 6G networks. &
TL + DRL. &
Reduces DRL training overhead by transferring knowledge from related tasks. &
TL is task-dependent; still relies on a reactive DRL core. &
Aids a reactive policy with TL; our method aims for broader generalization via distributional learning. \\
\hline

Chen \& Heydari \cite{chen2024resource} &
Resource governance via adaptive adjustment of network topology. &
VAE + DRL. &
Optimizes network structure using an RL agent in a VAE-compressed latent space. &
High training complexity; uses VAE for state compression, not generation. &
VAE is used for state representation for an RL agent; our method is a unified GAI for direct solution generation. \\
\hline

Qiao et al. \cite{qiao2025resource} &
Multi-timescale resource allocation for O-RAN slicing (radio \& computing). &
Parallel Hierarchical DRL. &
Decomposes the problem hierarchically to handle complexity and coupled constraints. &
High complexity of the hierarchical DRL architecture. &
Advanced DRL structure that still learns reactive policies; our method is a single, unified generative model. \\
\hline

Lotfi et al. \cite{lotfi2llm} &
Dynamic O-RAN slicing and resource management. &
LLM + MARL. &
Enriches RL state with semantic context from an LLM for improved planning. &
High computational cost of LLMs; relies on the quality of LLM output. &
Augments DRL state with LLM semantics; we learn robust features directly from raw data via contrastive learning. \\
\hline

Salama et al. \cite{salama2025fedora} &
Resource allocation in O-RAN specifically to support FL processes. &
FL + RL + Model-based. &
Optimizes O-RAN resources (e.g., power) to improve the efficiency of an ongoing FL task. &
Solves a different, specific problem (RA for FL), not a general network slicing problem. &
Different problem domains and uses a hybrid of non-generative methods. \\
\hline

Qazzaz et al. \cite{qazzaz2024machine} &
Dynamic PRB allocation in O-RAN based on traffic and QoS. &
Supervised ML (Random Forest Classifier). &
Provides fast decisions by selecting the best policy from a predefined, finite set of options. &
Cannot generate novel allocation policies; the quality of predefined policies bounds performance. &
A discriminative model that selects a policy; our model is generative and creates new allocation solutions. \\
\bottomrule
\end{tabularx}
\end{table*}

\section{System Model}
\label{system_model}

This study addresses the resource allocation problem in the \ac{O-RAN} architecture through network slicing, focusing on two primary slice types: \ac{URLLC} and \ac{eMBB}. The \ac{URLLC} slice is designed for ultra-low latency and high-reliability communications, which are essential for critical applications. In contrast, the \ac{eMBB} slice aims to deliver high data rates for mobile broadband services. Each slice type imposes distinct \ac{QoS} requirements on resource allocation, as conceptually illustrated for the \ac{O-RAN} functional split in \figref{fig:slicing} \cite{popovski20185g}. Based on \figref{fig:slicing}, the O-RU is shared across services. Within the O-DU, the High PHY layer is common, while the MAC and RLC layers are logically separated per slice. In the O-CU-UP, the SDAP and PDCP layers are also sliced to cater to differentiated service needs.

\begin{figure}[t]
	\centerline{\includegraphics[width=0.5\textwidth]{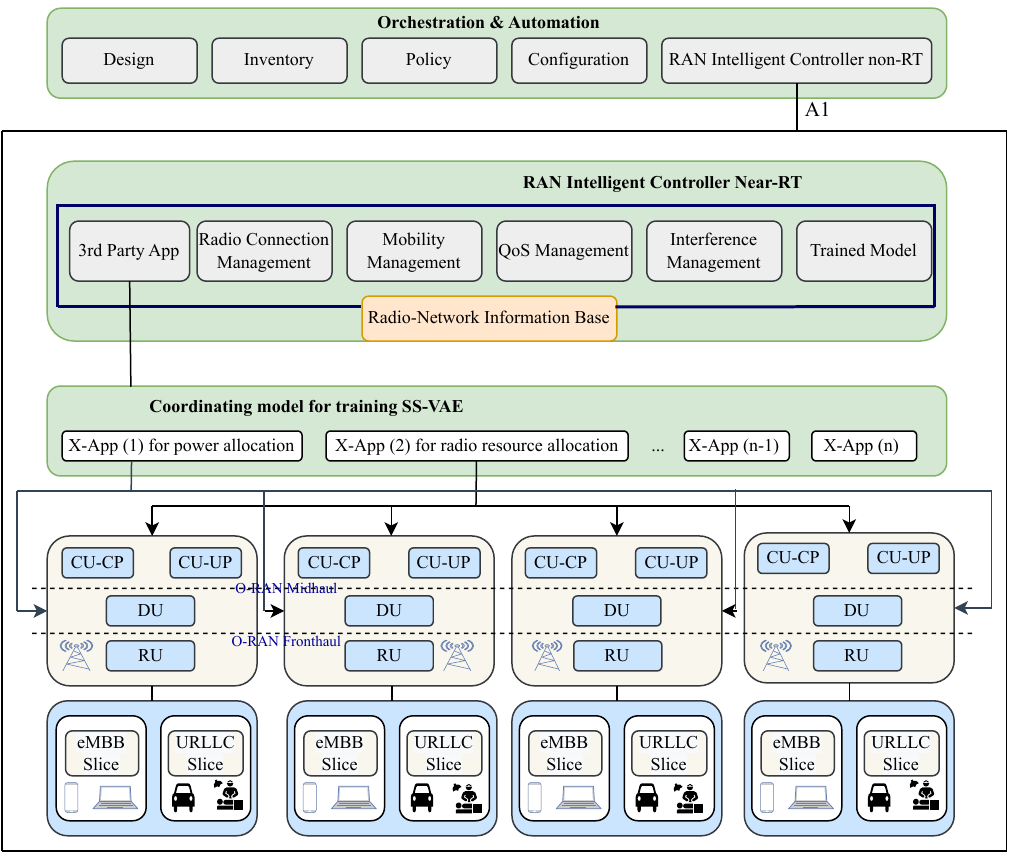}}
	\caption{\ac{O-RAN} network system model}
	\label{fig:oran system model}
\end{figure}

Logically slicing the \ac{RAN} into separate \ac{URLLC} and \ac{eMBB} service-aware components allows for optimized network operation tailored to diverse traffic characteristics, thereby enhancing overall resource efficiency and operational flexibility. Key resources considered for allocation in our \ac{O-RAN} model include transmission power and \acp{PRB}. Traffic demands are typically characterized by data rate and bandwidth requirements, while \ac{QoS} parameters specify the target performance metrics for each service. The overall system model is depicted in \figref{fig:oran system model}. All variables and parameters used throughout this paper are defined and detailed in \tabref{tab: notations}.

\begin{table}[t]
	\centering
	\begin{tabular}{p{1cm}|p{7cm}}
		\hline
		\textbf{Notation} & \textbf{Definition} \\ \hline
		$b, \mathcal{B}, B$ & Index, set, and the number of \acp{RU} \\ 
		$u, \mathcal{U}, U$ & Index, set, and the number of \acp{UE} \\ 
		$m, \mathcal{M}, M$ & Index, set, and the number of \acp{PRB} per \ac{RU} \\ 
		$s, \mathcal{S}, S$ & Index, set, and the number of slices \\
		$h, H$ & Index, and channel gain matrix between \acp{RU} and \acp{UE} \\ 
		\hline
		$\alpha_{u_s}^b$ & \ac{UE} $u$ association to \ac{RU} $b$ in slice $s$ \\
		$\beta_{u_s,m}^b$ & \ac{UE} $u$ association to \ac{PRB} $m$ of \ac{RU} $b$ in slice $s$ \\
		\hline
		$p_{u_s,m}^b$ & Transmission power of \ac{PRB} $m$ of \ac{RU} $b$ in slice $s$ \\
		$R_{u_s,m}^b$ & Rate achieved for \ac{UE} $u$ in \ac{PRB} $m$ of \ac{RU} $b$ in slice $s$ \\
		$R_{u_s}$ & Rate achieved for \ac{UE} $u$ in slice $s$ \\
		$R_{u}$ & Rate achieved for \ac{UE} $u$ \\
		$w_s$ & Priority weights for each service type \\
		$P^b$ & Transmitted power \ac{RU} in the fronthaul link \\
		$C^b$ & Data rate in the fronthaul link \\
		$\sigma_1^2$ & Quantization Noise \\
		\hline
		$D^{\text{pro}}$ & Overall propagation delay \\
		$D^{fr,p}$ & Propagation delay in the fronthaul link \\
		$D^{mid,p}$ & Propagation delay in the midhaul link \\
		$D^{bc,p}$ & Propagation delay in the backhaul link \\
		$L$ & Length of the fiber link \\
		$C$ & Propagation speed of the medium \\
		\hline
		$D^{tr}$ & Total transmission delay \\
		$D^{fr,t}$ & Transmission delay in the fronthaul link \\
		$D^{mid,t}$ & Transmission delay in the midhaul link \\
		$D^{bc,t}$ & Transmission delay in the backhaul link \\
		$R$ & Data rate of the packet \\
		$\mu$ & Mean packet size \\
		\hline
	\end{tabular}
	\caption{Notations with definitions for the system model parameters.}
	\label{tab: notations}
\end{table}

In our study, we explore a deployment based on \ac{O-RAN}, wherein each \ac{BS}, known as \ac{RU} in \ac{O-RAN}, supports two different slices: \ac{eMBB} and \ac{URLLC}. These slices are designed to meet the \ac{QoS} requirements of \acp{UE}. Our analysis focuses on two key network functions, power control and radio resource allocation, which are jointly considered.

We represent the set of all \acp{RU} as $\mathcal{B}$ and the set of \acp{UE} as $\mathcal{U}$. Each \ac{RU} $b$ serves two sets of \acp{UE}: $\mathcal{U}_{e}^b$ for \ac{eMBB} and $\mathcal{U}_{u}^b$ for \ac{URLLC}, leading to a total of $U$ \acp{UE}, where $U = U_{e}^b + U_{u}^b$. Each \ac{UE} is exclusively associated with a single \ac{RU}. The temporal and spectral layout is divided into $\mathcal{M}$ segments known as \acp{PRB}, which are the smallest unit of resources for allocation \cite{erpek2015optimal}. Each \ac{PRB} covers a specific time slot in the temporal domain and a frequency range of f Hz in the spectral domain. We denote $S_e$ for \ac{eMBB} slices and $S_u$ for \ac{URLLC} slices. 
% The total number of slices ($S$) will be $S = S_e + S_u$.

The \ac{UE} association is indicated by a binary variable $\alpha_{u_s}^b$ taking a value of 1 if \ac{UE} $u$ in slice $s$ is connected to \ac{RU} $b$  and 0 otherwise. Since each \ac{UE} should be connected to only one \ac{RU}, the following constraint must be met:
\begin{align}
	\sum_{b \in \mathcal{B}} \alpha_{u_s}^b = 1, \; \; \; \forall u \in \mathcal{U}, \; \; s \in \mathcal{S}.
\end{align}

We define the binary variable $\beta_{u_s,m}^b$ as the indicator for \ac{PRB} allocation, taking the value of 1 when \ac{UE} $u$ in slice $s$ is connected to \ac{PRB} $m$ of RU $b$, and 0 otherwise. We have
\begin{align}
	\beta_{u_s,m}^b \leq \alpha_{u_s}^b, \; \; \; \; \forall b \in \mathcal{B}, \; \; m \in \mathcal{M}, \; \; u \in \mathcal{U}, \; \; s \in \mathcal{S}.
\end{align}

Similarly, each \ac{PRB} $m$ of \ac{RU} $b$ is associated with only one \ac{UE}, as
\begin{align}
	\sum_{u \in \mathcal{U}} \alpha_{u_s}^b \beta_{u_s,m}^b \leq 1,   \; \; \; \; \forall b \in \mathcal{B}, m \in \mathcal{M}, \; \; s \in \mathcal{S}.
\end{align}

We denote the transmission power of \ac{PRB} $m$ in \ac{RU} $b$ in slice $s$ as $p_{u_s,m}^b$ for \ac{UE} $u$, and the channel gain between \ac{RU} $b$ and \ac{UE} $u$ in slice $s$ as $h_{u_s,m}^b$. The rate achieved for \ac{UE} \(u\) in slice \(s\) by allocating \ac{PRB} \(m\) of \ac{RU} \(b\) is denoted by \(R_{u_s,m}^b\), where \(\eta_{u_s,m}^b\) represents the \ac{SINR}. The calculation for \(R_{u_s,m}^b\) is determined by
\begin{align}
	& \eta_{u_s,m}^b = \frac{\alpha_{u_s}^b h_{u_s,m}^b p_{u_s,m}^b}
	{\sum_{j \neq b}^{N}\sum_{l \neq s}^S \sum_{i \neq u}^{U} \alpha_{i_l}^j h_{i_l,m}^j p_{i_l,m}^j + \sigma^2} , \\
	& R_{u_s,m}^b = \log_2(1 + \eta_{u_s,m}^b), \\
	& R_{u_s} = \sum_{b \in B} \sum_{m \in M} \alpha_{u_s}^b \beta_{u_s,m}^b R_{u_s,m}^b,\\
	%	& R_{u, b}(t) = \sum_{s \in \mathcal{S}} \sum_{m \in \mathcal{M}} \beta_{u_s,m}^b R_{u, b, m}(t)  \; \; \; \; \forall b \in \mathcal{B}, \; \; u \in \mathcal{U}.
	\label{eq: rate_u_s}
	R_u & = \sum_{s \in \mathcal{S}} \sum_{u \in \mathcal{U}} \alpha_{u, b, m, s} \sum_{m \in \mathcal{M}} \beta_{u_s,m}^b R_{u, b, m,s} \notag \\
	& = \sum_{u \in \mathcal{U}} \sum_{m \in \mathcal{M}} \beta_{u,m}^b R_{u_s,m}^b  \; \; \; \; \forall m \in \mathcal{M}, \; s \in \mathcal{S}.
\end{align}

The total power transmitted by \ac{RU} $b$ and the data rate for \ac{UE} in the fronthaul link are represented as $P^b$ and $C^b$, respectively. Quantization noise ($\sigma_1^2$) can potentially impact the signal's fidelity during transmission over the fronthaul, mainly when the original signal displays significant variations or spans a wide range of values. This noise may lead to inaccuracies in the transmitted signal, thereby influencing the \ac{QoS} perceived by the \ac{UE}. These values are determined as follows:
\begin{align}
	& P^b = \sum_{u=1}^U \sum_{m=1}^M \sum_{s=1}^S {\alpha_{u_s}^b h_{u_s,m}^b p_{u_s,m}^b} + \sigma_1^2 , \\
	& C^b = \log_2(1+ \frac{\sum_{u=1}^U \sum_{m=1}^M \sum_{s=1}^S {\alpha_{u_s}^b h_{u_s,m}^b p_{s,m}^s}}{\sigma_1^2}) \notag \\ 
	& = \log_2(\frac{P_b}{\sigma_1^2}).
\end{align}

The channel characteristics in our model are simulated using a Rayleigh fading process, which represents the variable channel conditions typical of \ac{NLOS} environments. The channel vector from \ac{PRB} $m$ in \ac{RU} $b$ to $u$ \ac{UE} in $s$ slice, denoted by $\mathbf{h}_{u_s,m}^b$, is modeled as $\mathbf{h}_{u_s,m}^b = \sqrt{\beta_{u_s,m}^b} \mathbf{g}_{u_s,m}^b$, where $\beta_{u_s,m}^b$ represents the large-scale fading coefficient, accounting for path loss and shadowing, and $\mathbf{g}_{u_s,m}^b \sim \mathcal{N}(0, N_0 I_{D_s})$ is the fast and flat fading effects modeled as a circularly symmetric complex Gaussian random variable. To reflect the impact of interference and uncertainties, the channel gain matrix $\mathbf{H}$ is augmented with Gaussian noise, expressed as $\mathbf{H}_{noise} = \mathbf{H} + \mathbf{\epsilon}$, where $\mathbf{\epsilon}$ is a Gaussian noise matrix with zero mean and adjustable variance $\sigma^2$. 

The total delay experienced by each \ac{UE} is the sum of propagation and transmission delays. The overall propagation delay ($D^{\text{pro}}$) consists of the delays in the fronthaul ($D^{fr,p}$), midhaul ($D^{mid,p}$), and backhaul ($D^{bc,p}$) links. In each of these links, the propagation delay is determined as the time it takes for a signal to traverse the distance, calculated as $D = L/C$, where $L$ is the link length and $C$ is the propagation speed. Similarly, the total transmission delay ($D^{tr}$) is the sum of transmission delays in the fronthaul ($D^{fr,t}$), midhaul ($D^{mid,t}$), and backhaul ($D^{bc,t}$) links. Within each link, the transmission delay corresponds to the time needed to transmit all packets into the transmission medium, determined by $D = \frac{\mu}{R_{u_s}}$, where $R_{u_s}$ is the data rate for \ac{UE} $u$ in slice $s$, and $\mu$ is the mean packet size. The total delay for \ac{UE} $u$ in slice $s$ is $D_{u_s} = L/C + \mu/R_{u_s}$, and should be below a specific threshold ($D^{\max}_{u_s}$), expressed as $D_{u_s} \leq D^{\max}_{u_s}$.

The objective of our resource allocation scheme is to optimize the overall network utility. This primarily involves maximizing the weighted sum of the data rates, $R_{u_s}$, achieved by all \acp{UE}, while simultaneously satisfying their respective \ac{QoS} requirements, particularly the delay constraints. The weight $w_s$ reflects the priority or importance associated with service type $s$ (e.g., \ac{eMBB} or \ac{URLLC}). The intelligent resource allocation function, managed by \acp{xApp} at the \ac{RIC} and influencing \ac{RU} behavior, is responsible for effectively managing these diverse requirements. Hence, we can mathematically express the optimization problem as follows:

\begin{align}  
	\label{eq: opt problem ra}
	\max_{\mathbf{p,\alpha, \beta}}  & \sum_{s \in \mathcal{S}}\sum_{u \in \mathcal{U}}  w_s R_{u_s}  \\
	\text{subject to} \notag \\
    \displaybreak[1]
	\label{eq:power_limit}
	& P^b \leq P_{b}^{max} \;\;\; \forall b \in \mathcal{B} \\
    \displaybreak[1]
	\label{eq:nonnegativity}
	& p_{u_s,m}^b \geq 0 \;\;\; \forall b \in \mathcal{B}, m \in \mathcal{M}, s \in \mathcal{S}, u \in \mathcal{U} \\
    \displaybreak[1]
	\label{eq:max_power_per_stream}
	& p_{u_s,m}^b \leq P_{s}^{max} \;\;\; \forall b \in \mathcal{B}, m \in \mathcal{M}, s \in \mathcal{S}, u \in \mathcal{U} \\
    \displaybreak[1]
	\label{eq:min_rate}
	& R_{u_s} \geq R_{s}^{min} \;\;\; \forall u \in \mathcal{U}, s \in \mathcal{S} \\
    \displaybreak[1]
	\label{eq:capacity_limit}
	& C^b \leq C_{b}^{max} \;\;\; \forall b \in \mathcal{B} \\
    \displaybreak[1]
	\label{eq:delay_constraint}
	& D_{u_s} \leq D^{max}_{u_s} \; \; \; \forall u \in \mathcal{U}, s \in \mathcal{S} \\
    \displaybreak[1]
	\label{eq:alpha_binary}
	& \alpha_{u_s}^b \in \{0,1\} \;\;\; \forall u \in \mathcal{U}, b \in \mathcal{B}, s \in \mathcal{S} \\
    \displaybreak[1]
	\label{eq:beta_binary}
	& \beta_{u_s,m}^b \in \{0,1\} \;\;\; \forall u \in \mathcal{U}, b \in \mathcal{B}, m \in \mathcal{M}, s \in \mathcal{S} \\
    \displaybreak[1]
	\label{eq:unique_bs_assignment}
	& \sum_{b \in \mathcal{B}} \alpha_{u_s}^b = 1 \; \; \; \forall u \in \mathcal{U}, \; \; s \in \mathcal{S} \\
    \displaybreak[1]
	\label{eq:beta_alpha_dependency}
	& \beta_{u_s,m}^b \leq \alpha_{u_s}^b \; \; \; \; \forall b \in \mathcal{B}, m \in \mathcal{M}, u \in \mathcal{U}, s \in \mathcal{S} \\
	\label{eq:subcarrier_exclusivity}
	& \sum_{u \in \mathcal{U}} \alpha_{u_s}^b \beta_{u_s,m}^b \leq 1   \; \; \; \; \forall b \in \mathcal{B}, m \in \mathcal{M}, s \in \mathcal{S}
\end{align}

Equation \eqref{eq: opt problem ra} defines the objective of maximizing the total weighted data rate across all users and services. Constraints \eqref{eq:power_limit}, \eqref{eq:nonnegativity}, and \eqref{eq:max_power_per_stream} enforce limitations on the transmit power for each \ac{RU}, ensure non-negative power allocation for each \ac{UE}, and restrict the received power per \ac{UE}, respectively.

Constraint \eqref{eq:min_rate} guarantees the \ac{QoS} by ensuring that each service meets its minimum required data rate. Furthermore, \eqref{eq:capacity_limit} and \eqref{eq:delay_constraint} account for the fronthaul capacity limitations of each \ac{RU} and the delay constraints for each \ac{UE}, respectively, by their \ac{QoS} requirements.

Constraints \eqref{eq:alpha_binary} and \eqref{eq:beta_binary} define the binary nature of the decision variables $\alpha$ and $\beta$. Finally, constraints \eqref{eq:unique_bs_assignment}, \eqref{eq:beta_alpha_dependency}, and \eqref{eq:subcarrier_exclusivity} ensure that each \ac{UE} is associated with exactly one \ac{RU} and that \ac{RB} assignments are only made if the \ac{UE} is associated with the corresponding \ac{RU} while preventing multiple \acp{UE} from sharing the same \ac{RB} on the same \ac{RU}.
 
This formulation can be seen as a well-known NP-hard problem, a multi-dimensional knapsack problem. Since we can reduce it to our resource allocation problem in polynomial time, it follows that it is also NP-hard. Given the NP-hard nature of the resource allocation problem, finding an exact solution in polynomial time is computationally infeasible for large-scale networks.

The system model is constrained by limited resources, such as \ac{UE} or \ac{RU}, and slice power and fronthaul capacity, which prevents the aggregate throughput from exceeding its optimal value. This ensures that the objective function, representing aggregate throughput, has an upper bound and cannot increase infinitely. As a result, the algorithm used to solve the system will converge to an optimal solution. If the objective function is a strictly increasing function concerning the number of iterations, it will converge to its global optimum. The algorithm will converge to a local optimum if the function is non-monotonically ascending. This guarantees the stability and convergence of the system model within the feasible region.
\section{Methodology}
\label{Proposed_scheme}

To address the complex resource allocation problem in the \ac{O-RAN} architecture, we introduce SS-VAE, a novel \ac{GAI} framework rooted in semi-supervised learning. This method employs a specialized \ac{DNN} architecture that intrinsically integrates a \ac{VAE} —leveraging its ability to learn underlying data distributions—with a contrastive loss function designed to enhance representation quality. The SS-VAE is designed to comprehend complex relationships between input features and optimal resource allocation decisions (e.g., parameters $\alpha$, $\beta$, and transmission powers) that meet stringent \ac{QoS} requirements. A comprehensive dataset underpins the development and initial supervised training of SS-VAE. This dataset features diverse \ac{O-RAN} scenarios, with labeled data representing optimal allocation parameters generated via an \ac{ESA}, which also serves as an optimal performance benchmark.

\begin{figure*}[t]
	\centering
	\centerline{\includegraphics[width=0.9\textwidth]{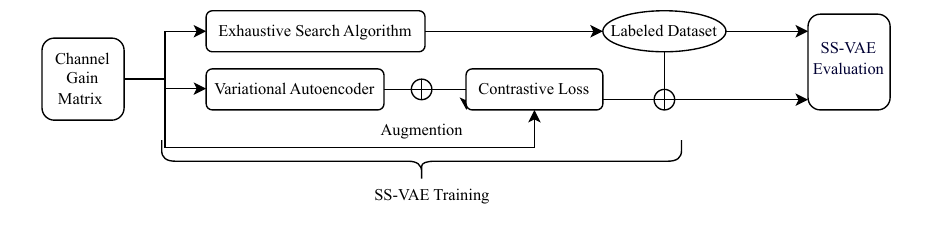}}
	\caption{Overview of the proposed methodology.}
	\label{fig:model overview}
\end{figure*}

The robustness and efficacy of the SS-VAE framework will be rigorously evaluated using multiple regression-based metrics suitable for assessing the accuracy of the generated resource allocation parameters. Its performance will be benchmarked against both the globally optimal solutions derived from \ac{ESA} and deployment-ready, state-of-the-art methods such as \ac{DRL} agents. This comprehensive validation aims to demonstrate the reliability of SS-VAE across various operational scenarios, particularly highlighting its capability to approach optimal performance without the prohibitive computational cost associated with \ac{ESA}.
\figref{fig:model overview} illustrates our proposed methodology, which encompasses the training pipeline for SS-VAE and a comparative evaluation against benchmark algorithms.

% --------------------------------------------------------------------------------------------------
\subsection{Exhaustive Search Algorithm (ESA) }

We employ \ac{ESA} method as a basis of comparison for our algorithm. The \ac{ESA} is a fundamental method to find the global optimal solution for resource allocation problems by exhaustively evaluating all possible configurations based on a predefined objective function \cite{eiselt2019nonlinear}. It serves as a benchmark by providing a ground truth for comparing other methods. While it is simple to understand and guarantees optimality within a finite time, its high computational cost limits its use to small or well-defined problems \cite{nocedal1999numerical}. Nevertheless, it remains a widely used benchmark for evaluating and comparing optimization algorithms.

We use the \ac{ESA} to create a labeled dataset by obtaining a subset of optimal solutions for the objective variables (i.e., $\alpha$, $\beta$, $p$) in the resource allocation optimization problem. This dataset is subsequently used to train the supervised learning component of our SS-VAE method. The detailed steps of the \ac{ESA} are described in Algorithm \ref{alg:esa}.

\begin{algorithm}
	\caption{ESA for Resource Allocation in O-RAN}
	\label{alg:esa}
	\begin{algorithmic}[1]
		\Require $B$, $U$, $S$, $M$, $w_s$ , $P^{max}_b$, $P^{max}_s$, $R^{min}_s$, $C_b^{max}$, $D_{u,s}^{min}$, and $H$
		
		\Ensure Optimal resource allocation $best\_solution$ and its objective value $best\_value$
		
		\State Initialize dataset $\mathcal{X} \gets \emptyset$
		
		\State $best\_value \gets -\infty$
		\State $best\_solution \gets \emptyset$
		
		\For{each possible combination of $(\alpha, \beta, p)$}
		\State Calculate $R_{u,s}$ using Eq. \eqref{eq: rate_u_s}
		\If{all constraints are satisfied}
		\State $current\_value \gets \sum_{s \in S} \sum_{u \in U} w_s R_{u,s}$
		\If{$current\_value > best\_value$}
		\State $best\_value \gets current\_value$
		\State $best\_solution \gets (\alpha, \beta, p)$
		\State Add $(\alpha, \beta, p, R_{u,s})$ to dataset $\mathcal{X}$
		\EndIf
		\EndIf
		\EndFor
		
		\State \Return $best\_solution$, $best\_value$, $\mathcal{X}$
		
	\end{algorithmic}
\end{algorithm}

% --------------------------------------------------------------------------------------------------

\subsection{Deep Reinforcement learning Algorithm}
\label{sec: DRL}

To evaluate the proposed SS-VAE methodology, we benchmark its performance against a \ac{DRL} model, a paradigm increasingly recognized for its efficacy in navigating complex resource allocation tasks within contemporary wireless networks \cite{filali2023communication}. The resource allocation challenge, as formulated in Problem \ref{eq: opt problem ra}, is cast as a \ac{MDP}. This formalization allows the \ac{O-RAN} orchestrator, responsible for \ac{PRB} assignments and \ac{RU} associations per slice, to operate as a learning agent \cite{filali2023communication, feriani2021single, suh2022deep, fan2020theoretical}. The \ac{MDP} is characterized by: 

\begin{enumerate}[label=\Roman*.]
	\item \textbf{State}: The state at time \(t\) (\(\mathfrak{s}(t) \in \mathfrak{S}\)), is represented by \(\{\mathfrak{s}_u(t)\}_{t=1}^{N}\), where \(\mathfrak{s}_u(t)\) indicates the state of \ac{UE} \(u\) (\(u=1:U\)). Here, \(\mathfrak{s}_u(t)\) is a binary indicator: it equals one if the data rate requirement of \ac{UE} \(u\) is satisfied and 0 otherwise. Each data rate also corresponds to a quantized transmission power level.
	\item \textbf{Action}: An action (\(\mathfrak{a} \in \mathfrak{A}\)), is represented as $\mathfrak{a} = \{ \alpha_{u, b, s}, \{\beta_{u,b,m,s}\}_{m=1}^{\mathcal{M}} \}_{b=1}^{\mathcal{B}}$.
	\item \textbf{Transition Probability}: 
	The transition probability \(\mathfrak{P}(\mathfrak{s} | \mathfrak{s}, \mathfrak{a})\) represents the probability of moving from one state \(\mathfrak{s}\) to another one \(\mathfrak{s}^\prime\) by taking action \(\mathfrak{a}\). It is influenced by several factors, including the \ac{QoS} requirements of \acp{UE} and the resource allocation decisions made for \acp{UE}.
	
	\item \textbf{Reward}: The reward function evaluates the gains or costs of a given state-action pair. It guides the agent's decision-making process by indicating the desirability of various actions across different states. Consistent with previous studies \cite{filali2023communication, feriani2021single, suh2022deep}, our approach defines the reward function as a combination of the objective function and the constraints pertinent to the considered problem (Equation \ref{eq: opt problem ra}). Thus, we specify the reward function for our specific scenario as follows:
	\begin{align}
		& \mathfrak{R(s, a)} = \Theta_r \; R_{u_s} + \Theta_{const.} \; C_{u_s, m}^b + \Theta_{bias},
	\end{align}
	where, \( \Theta_r \), \( \Theta_{const.} \), and \( \Theta_{bias} \) represent the respective weights assigned to the objective function (Data Rate), constraints, and the bias value. The primary objective of the agent is to determine the optimal policy, denoted as \( \pi: \mathfrak{S} \rightarrow \mathfrak{P(A)} \), for the \ac{MDP}, that maximizes the cumulative reward obtained through dynamic learning from acquired data.
\end{enumerate}

The inherent complexity of wireless resource allocation, characterized by high-dimensional state and action spaces and continuous variables (e.g., transmission power), renders traditional Q-learning approaches inefficient due to the curse of dimensionality and issues with generalizing from sparsely visited state-action pairs \cite{filali2023communication, fan2020theoretical}. We employ a \ac{DQN} framework to surmount these challenges. \ac{DQN} leverages a \ac{DNN} to approximate the action-value function (Q-function), enabling it to manage large-scale problems effectively. Crucially, \ac{DQN} incorporates mechanisms such as experience replay—storing transitions ($\mathfrak{S}_t, \mathfrak{A}_t, \mathfrak{R}_t, \mathfrak{S}_{t+1}$) in a replay memory ($\mathbb{M}$) to break data correlations—and target networks to stabilize the learning process and improve convergence.

To ensure a robust exploration-exploitation balance during training, the \ac{DQN} agent utilizes an epsilon-greedy strategy for action selection:
\[ \pi(S) = \begin{cases} \text{argmax}_\mathfrak{a} Q(\mathfrak{S, a}) & \text{with probability } 1 - \epsilon \\ \text{random action} & \text{with probability } \epsilon \end{cases} \]

This allows the agent to exploit its current knowledge by selecting the action with the highest estimated Q-value with probability ($1-\epsilon$) or to explore the action space randomly with probability ($\epsilon$). The comprehensive training algorithm for the \ac{DQN} is detailed, and our specific implementation for this problem is encapsulated in Algorithm 3 \cite{fan2020theoretical}.

% -----------------------------------------------------------------------------------------------------------------------
\subsection{Semi-Supervised \ac{VAE} (SS-VAE)}
% {\color{red} this section should start with your method and should not have introduction.}
% Traditional resource allocation methods often struggle with dynamic network conditions. \ac{DL} techniques offer improved accuracy and efficiency by learning complex, non-linear relationships between inputs (e.g., channel matrices) and outputs (e.g., resource allocation decisions). \ac{DL}-based approaches adapt to changing network conditions, user demands, and interference, offering more flexible and efficient resource management than traditional methods, which are hampered by predefined rules and a lack of real-time learning \cite{lima2020resource, iturria2022multi, hammami2022policy, wang2023resource}.

To address the resource allocation problem, we introduce our SS-VAE method in this subsection. The model is designed to process a flattened channel gain matrix (dimension $U \times B$) through a \ac{DNN} to find optimal resource allocation strategies. The central contribution of our method is the integration of a \ac{VAE} with a contrastive loss, which improves the learning of refined representations..

In our framework, \acp{xApp} for power transmission and \ac{PRB} allocation coordinate through a shared \ac{VAE} foundation. This \ac{VAE} acts as a common space, allowing xApps to collaboratively consider relevant features by accessing a shared latent representation of channel gain matrices. This shared understanding facilitates a cohesive analysis and simultaneous consideration of key features impacting both domains. This leads to more comprehensive and informed resource allocation decisions and enables synchronized \ac{O-RAN} resource management.

\subsubsection{Variational Autoencoder}
\begin{figure*}[t!]
	\centerline{\includegraphics[width=0.8\textwidth]{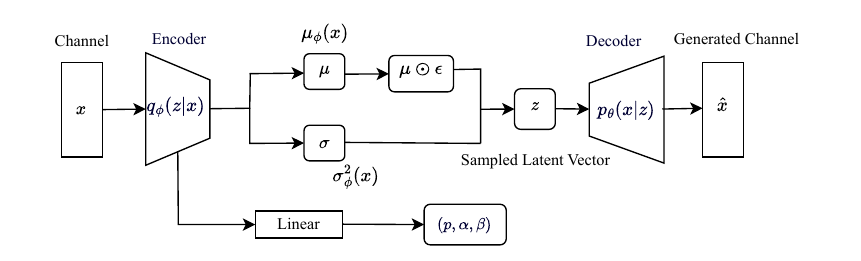}}
	\caption{Architecture of the \ac{VAE}.}%\color{red} fonts are too small.}
	\label{fig:vae arch}
\end{figure*}

The \ac{VAE}, a \ac{DNN} generative model for unsupervised learning, excels at capturing the intrinsic distribution of input data like the channel gain matrix, which is valuable for the resource allocation problem \cite{kingma2019introduction}. It maps input data $x$ (flattened channel gain matrix) to a lower-dimensional latent space $z$, with $p_\theta(x)$ as the generative model. The \ac{VAE} aims to maximize data likelihood given model parameters $\theta$, as expressed mathematically by:
\begin{align}
	\log p_\theta(x) = \mathbb{E}{q_\phi(z|x)}[\log p_\theta(x|z)] - D_{KL}(q_\phi(z|x) | p(z)),
	\label{eq: vae loss}
\end{align}

\noindent where $q_\phi(z|x)$ is the encoder to model the posterior distribution of the latent variable based on the input data, and $D_{KL}(q_\phi(z|x) | p(z)), $ is the \ac{KL} divergence this posterior distribution and the prior distribution of the latent variable. Optimizing the loss (Equation \ref{eq: vae loss}) involves adjusting generative model ($\theta$) and encoder ($\phi$) parameters \cite{kingma2019introduction}. As illustrated in \figref{fig:vae arch}, the \ac{VAE} 's encoder compresses the channel gain matrix into a lower-dimensional latent code, and the decoder reconstructs it, minimizing reconstruction loss during training.

Using \acp{VAE} for resource allocation offers advantages like representing complex distributions, capturing non-linear relationships, and providing a compressed latent space useful for other tasks. They offer a versatile solution to inherently non-convex and challenging resource allocation problems.

The dataset used to train the SS-VAE model consists of $Q$ training samples, divided into two categories: $Q_L$ labeled samples and $Q_U$ unlabeled samples. These samples are represented by pairs of channel matrices and their corresponding power allocation vectors, denoted as $H_i$ and $\Gamma_i$ for labeled samples (where $i$ ranges from 1 to $Q_L$), and $H_j$ for unlabeled samples (where $j$ ranges from 1 to $Q_U$). The labeled dataset results from the ESA applied to the same training dataset. Specifically, we have used power transmission and allocation decisions ($p, \alpha, \beta$) obtained by ESA as labels for training our SS-VAE model. The supervised learning component of our algorithm uses the ADAM \cite{kingma2014adam} optimizer and an \ac{MSE} loss function for the $Q_L$ labeled samples, defined as follows:
\begin{equation}
	L_{supervised} = \frac{1}{\beta_L} \sum_{i \in \beta_L} || encoder(VAE(H_i)) - \Gamma_i||^2.
\end{equation}

The encoder part of the \ac{VAE} is used to estimate the transmission power and predict the allocation indicators ($\alpha_{u_s}^b$ and $\beta_{u_s,m}^b$) from the flattened $UB$ dimensional input channel gain matrix.

\subsubsection{Contrastive Loss}
Contrastive loss, common in \ac{DL} for representation learning, aims to map similar data points closer and dissimilar ones further apart in the representation space \cite{chen2020simple}. Mathematically, it can be expressed as:
\begin{align} 
	& L = \frac{1}{2N} \sum_{i=1}^{N} ( y_i \cdot d^2(\mathbf{x}_i, \mathbf{x}_i^+) \notag \\
	& + (1-y_i) \cdot \max(0, m-d^2(\mathbf{x}_i, \mathbf{x}_i^-)) ) ,
\end{align}

\noindent where $N$ is the number of data points, $\mathbf{x}_i$ is the $i^{th}$ data point, $\mathbf{x}_i^+$ and $\mathbf{x}_i^-$ are similar and dissimilar data points, $y_i$ is a binary label indicating whether $\mathbf{x}_i$ and $\mathbf{x}_i^+$ are similar or dissimilar,  $d(\mathbf{x}_i, \mathbf{x}_i^+)$ is the distance function between $\mathbf{x}_i$ and $\mathbf{x}_i^+$, and $m$ is a margin that separates the similar and dissimilar data points.

To apply contrastive loss in our resource allocation problem, we generate similar and dissimilar channel gain matrices using a channel statistical model that captures the communication environment. Introducing randomness via a Rayleigh fading model produces matrices with varying magnitude and phase coefficients. Specifically, we create a complex Gaussian random matrix with zero mean, where the variance depends on transmitter-receiver distance and accounts for interference as noise \cite{geng2015optimality, naderializadeh2014itlinq}. Adjusting this variance controls channel correlation, creating similar or dissimilar matrices. With these, we train our model using the contrastive loss function formulated as follows:
\begin{align}
	& L_{Contrastive}(\theta) = \frac{1}{\beta_L} \notag \\  
	& \sum_{i \in \beta_L}  [-\log \frac{ \exp^{ \frac{ f(\bar{H_i})^T \times f(\underline{H_i})} {\tau}}}
	{\exp^{ \frac{ f(\bar{H_i})^T \times f(\underline{H_i})} {\tau}} + 
		\sum_{i \neq j} \exp^{ \frac{ f(\bar{H_i})^T \times f(\underline{H_j})} {\tau}}}] ,
\end{align}

\noindent where $\theta$ is the set of parameters in the \ac{VAE} model, $\tau$ is a temperature hyperparameter, and $L(\theta)$ is the contrastive loss function. $\underline{H_i}$ and $\bar{H_i}$ are matrices similar to the main channel gain matrix $H_i$, used for training the SS-VAE model \cite{naderializadeh2021contrastive}. Minimizing this loss finds $\theta$ that enables the model to distinguish similar and dissimilar channel gain matrices, thereby addressing the resource allocation problem. The connection between the \ac{VAE} and the contrastive loss is key for optimal performance. The first step uses supervised learning to learn a good representation of the problem, and the next step refines these representations, improving the model's generalization to new, unseen scenarios.

\subsubsection{Training the SS-VAE model}
The SS-VAE training consists of two parts: supervised learning (i.e., training the \ac{VAE} using its encoder for prediction) and unsupervised learning (i.e., utilizing contrastive loss to enhance generalization and robustness). Key hyperparameters include the latent space dimension (balancing expressiveness and overfitting) \cite{shi2011iteratively}, learning rate, batch size, and the number of epochs. $L1$ and $L2$ regularization mitigates overfitting \cite{janocha2017loss}. The model uses PyTorch \cite{paszke2019pytorch}, with a 20\% validation/test split. Tuned hyperparameters are in \tabref{tab:param_training}.

\subsubsection{Performance Metrics}
The performance metrics for evaluating our approach include Mean Absolute Error (MAE) ($\text{MAE} = 1/N \sum_{i=1}^N |y_i - \hat{y}_i| $), cosine similarity ($\text{Cosine-Similarity}(Y, \hat{Y}) = Y \cdot \hat{Y}/ ( |Y| |\hat{Y}|)$), and Pearson correlation ($\text{correlation}(\hat{Y}, Y) = \text{cov}(\hat{Y}, Y)/ (\sigma_{\hat{Y}} \sigma_{Y})$). Here, \(N\) is the total number of samples, \(y_i\) and \(\hat{y}_i\) represent the true and predicted values, the dot product is denoted by \(\cdot\), and vector magnitude by \(| \cdot |\). The covariance between predicted and actual values is \(\text{cov}(\hat{Y}, Y)\), with \(\sigma_{\hat{Y}}\) and \(\sigma_{Y}\) as their standard deviations.
\section{SIMULATION RESULTS}
\label{simulation_results}

We analyzed the initial values in the following subsections to simulate the SS-VAE algorithm. Subsequently, we will compare the results obtained using SS-VAE with those from \ac{ESA} and \ac{DQN}.

\subsection{Simulation Configuration}

This section presents the numerical results for the resource allocation problem to evaluate the performance of the SS-VAE model compared to \ac{ESA} and \ac{DQN}. In our training and testing scenarios, each \ac{RU} in the \ac{O-RAN} deployment is equipped with 25 \acp{PRB} and operates with a single antenna with 10 UEs for each service. We consider a network with two slices/services, \ac{eMBB} and \ac{URLLC} services. These UEs are randomly distributed within a single-cell environment, mostly 10 to 500 meters from the \ac{RU}. The initial values and parameters used to run and simulate SS-VAE and the benchmark algorithms (\ac{ESA} and \ac{DQN}) are provided in \tabref{tab: param sim}, \tabref{tab: param dqn}, and \tabref{tab:param_training}.

For traffic generation, URLLC packets follow a constant bit rate (CBR) model with small fixed payloads of 32 bytes transmitted every 1 ms, resulting in a data rate of approximately 256 kbps per URLLC user. In contrast, eMBB traffic is modeled using a Poisson arrival process with packets sized at 1000 bytes to emulate realistic bursty user behavior such as web browsing and file downloads. 

For the channel modeling, we adopt the standard 3GPP-compliant formulation where the channel coefficient between user $u_s$ and RU $b$ on PRB $m$ is given by $ \mathbf{h}_{u_s,m}^b = \sqrt{\beta_{u_s,m}^b} , \mathbf{g}_{u_s,m}^b $. The large-scale fading component $\beta_{u_s,m}^b$ accounts for both path loss and shadowing and is computed as $ \beta_{u_s,m}^b = 10^{-(\text{PL}_{\text{dB}} + S)/10} $,
where the path loss in dB is modeled using the 3GPP Urban Macro (UMa) path loss model which is formulated as $\text{PL}_{\text{dB}} = 13.54 + 39.08 \log_{10}(d) + 20 \log_{10}(f_c)$, where $d$ is the 2D distance between the UE and RU in meters, and $f_c$ is the carrier frequency in GHz. For $f_c = 2$ GHz, this simplifies to
$\text{PL}_{\text{dB}} = 13.54 + 39.08 \log_{10}(d) + 6.02.$
Shadow fading $S$ follows a log-normal distribution $ \mathcal{N}(0, 8^2) $.

The small-scale fading component $ \mathbf{g}_{u_s,m}^b $ is modeled as a Rayleigh fading coefficient, i.e., $ \mathbf{g}_{u_s,m}^b \sim \mathcal{CN}(0,1) $, to represent non-line-of-sight propagation conditions.

\begin{table*}[t]
	\centering
	\begin{minipage}{0.5\textwidth}
		\centering
		\caption{Resource Allocation Simulation Parameters}
		\label{tab: param sim}
		\begin{tabular}{|c | c|} 
			\hline
			\textbf{Parameter} & \textbf{Value} \\ [0.5ex] 
			\hline
			Noise power & -174 dBm \\
			\hline
			bandwidth of each sub-carrier & 180 KHz \\
			\hline
			Maximum transmit power of each O-RU &  40 dBm \\
			\hline
			Maximum fronthaul capacity &  46 bps/Hz \\
			\hline
			Maximum data rate for eMBB &  10 bps/Hz \\
			\hline
			Maximum data rate for URLLC & 2 bps/Hz \\
			\hline
			Maximum power for URLLC \& eMBB & 30 dBm \\ [1ex] 
			\hline
		\end{tabular}
		\caption{DQN Training Parameters}
		\begin{tabular}{|c | c|}
			\hline
			Number of episodes and steps & 250, 50 \\
			\hline
			$\gamma$, $\epsilon_{\text{final value}}$, $\epsilon_{\text{decay factor}}$ & 0.9, 0.1, 0.9995 \\
			\hline
			Replay memory size & 400 \\
			\hline
		\end{tabular}
		\label{tab: param dqn}
	\end{minipage}%
	\hfill
	\begin{minipage}{0.5\textwidth}
		\centering
		\caption{SS-VAE Model Training Parameters}
		\label{tab:param_training}
		\begin{tabular}{ |c|c|c| } 
			\hline
			\textbf{Type} & \textbf{Training Parameter} & \textbf{Value} \\ [0.5ex] 
			\hline\hline
			Optimizer & Initial learning rate & 0.001 \\ 
			& $\beta_1$ & 0.99 \\ 
			& $\beta_2$ & 0.99 \\ 
			& Weight Decay & 0.9 \\
			\hline
			Model & Number of encoding layer & 20 \\
			& Dimension of latent space & 20 \\
			& Contrastive loss temperature & 0.25 \\
			& Activation Function & ReLU \\
			& Dropout Rate & 0.3 \\
			\hline
			Training & Epoch & 40 \\
			& Batch Size & 128 \\
			& Train/Validation Split & 0.2 \\ [1ex]
			\hline
		\end{tabular}
	\end{minipage}
\end{table*}

\subsection{Performance of Model Training}

\begin{figure}[t]
	\centerline{\includegraphics[width=0.95\columnwidth]{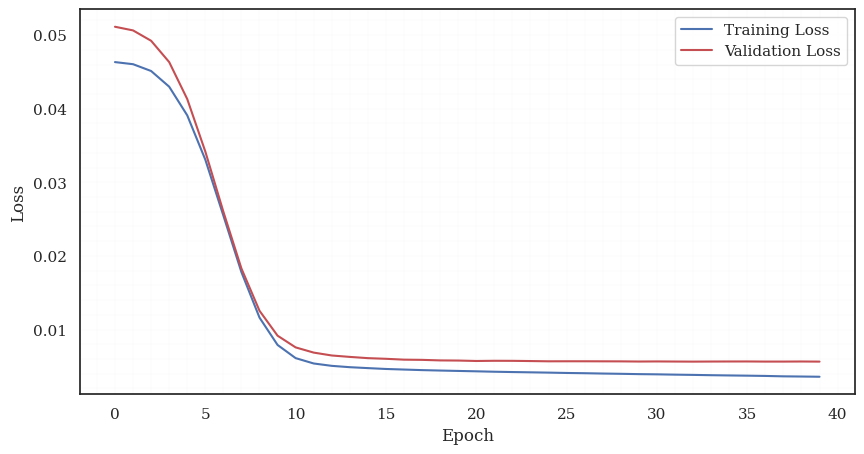}}
	\caption{Loss values of training and validation dataset.}
	\label{fig:loss_train_val}
\end{figure}

This section describes the performance of the trained model. \figref{fig:loss_train_val} shows the training and validation loss values. The decreasing trend of loss values indicates that the SS-VAE method could accurately estimate the target values (power of transmission, association indicators). The model can map the input channel gain matrix to the power of transmission and the indicators by employing the encoder part of the \ac{VAE} to learn the underlying information.

The performance metrics for the test data, as given in \tabref{tab:perf metrics}, clearly indicate that the SS-VAE method obtained optimal performance in the test dataset, demonstrating its generalization ability. The $R^2_{Score}$ value for the test data indicates that our model successfully explains a significant portion of the variance in estimating the power of transmissions. This improvement can be attributed to the utilization of \ac{VAE}, which has been proven to enhance generalization in various related tasks. Furthermore, the pre-training method involving contrastive loss has improved our model's ability to learn the underlying features, reducing generalization errors. 

\begin{table}[t]
	\centering
	\caption{Performance metrics of the SS-VAE model}
	\begin{tabular}{ |c | c | c | c | }
		\hline
		\textbf{Performance Metric} & \textbf{MAE} & \textbf{$R^2_{\text{Score}}$} & \textbf{Cosine Similarity} \\
		\hline
		Value & 0.09207532 & 0.76795106 & 0.988023 \\ 
		\hline
	\end{tabular}
	\label{tab:perf metrics}
\end{table}

\subsection{Comparison of the Algorithms}
\begin{figure*}[t]
  \centering

  % Row 1
  \subfloat[Aggregated throughput of SS-VAE and ESA.]{
    \includegraphics[width=0.45\textwidth]{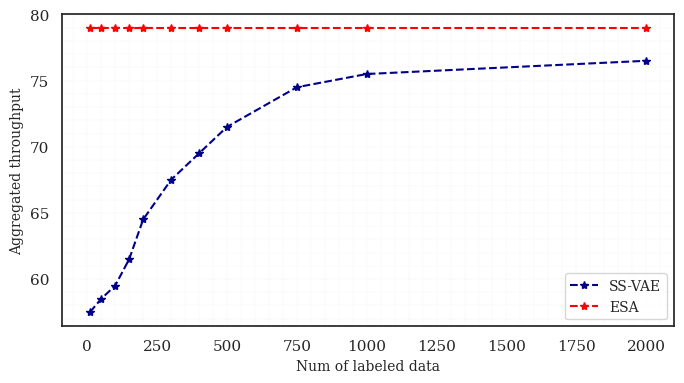}
    \label{fig:cmp_alg}
  }
  \hfill
  \subfloat[Association error of SS-VAE and ESA.]{
    \includegraphics[width=0.45\textwidth]{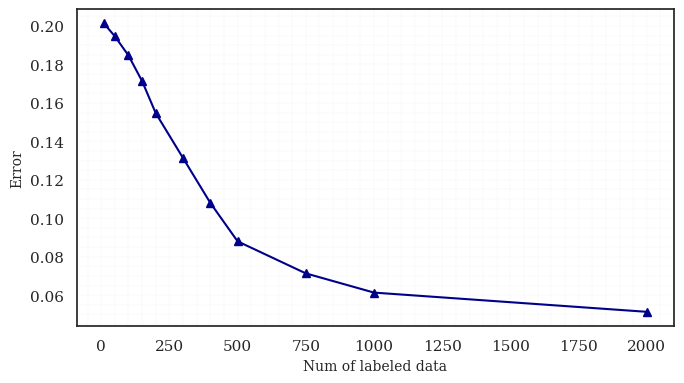}
    \label{fig:sum_diff_bin_vars}
  }

  \vspace{1em}

  % Row 2
  \subfloat[Aggregated throughput vs. number of UEs.]{
    \includegraphics[width=0.45\textwidth]{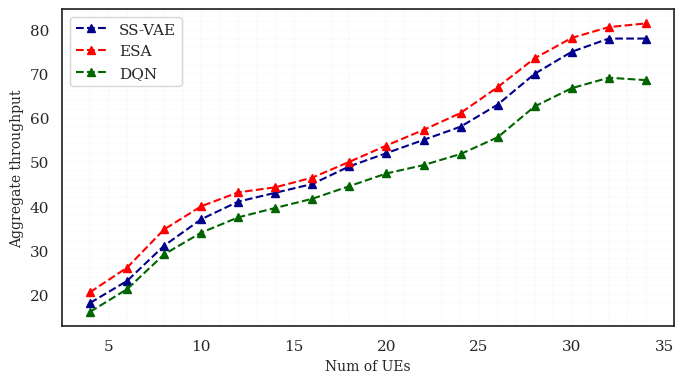}
    \label{fig:agg_rate_ue}
  }
  \hfill
  \subfloat[Aggregated throughput vs. number of UEs ($P_{O\text{-}RU}$).]{
    \includegraphics[width=0.45\textwidth]{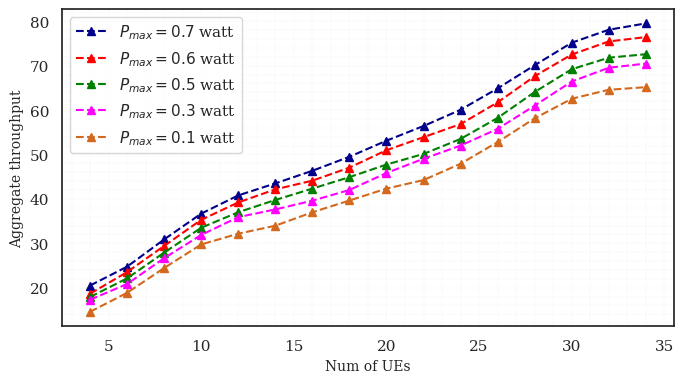}
    \label{fig:p_ru_agg_rate}
  }

  \vspace{1em}

  % Row 3
  \subfloat[Aggregated throughput vs. O-RU power.]{
    \includegraphics[width=0.45\textwidth]{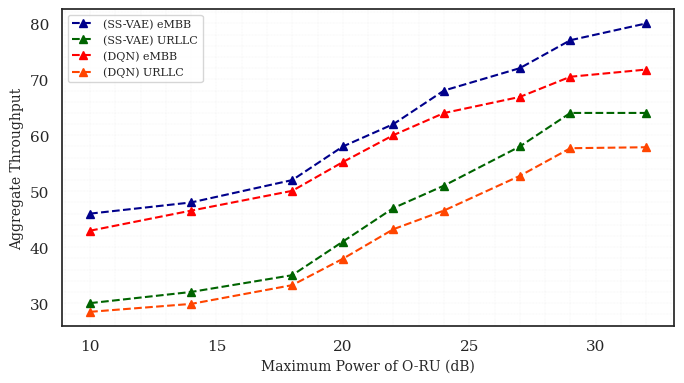}
    \label{fig:urrllc_embb_p_ru}
  }
  \hfill
  \subfloat[Aggregated throughput vs. Power of slice.]{
    \includegraphics[width=0.45\textwidth]{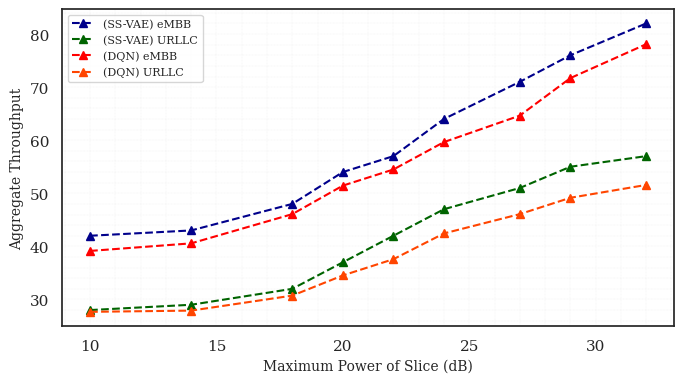}
    \label{fig:eu_p_slice}
  }

  \caption{Simulation results showing various performance metrics and algorithm comparisons for SS-VAE, ESA, and DQN models.}
  \label{fig:simulation_grid}
\end{figure*}
% \begin{figure*}[t]
%   \centering
%   \subfloat[Aggregated throughput of SS-VAE and ESA.]{
%     \includegraphics[width=0.3\textwidth]{figs/cmp_alg.png}
%     \label{fig:cmp_alg}
%   }
%   \quad
%   \subfloat[Association error of SS-VAE and ESA.]{
%     \includegraphics[width=0.3\textwidth]{figs/sum_diff_bin_variables.png}
%     \label{fig:sum_diff_bin_vars}
%   }
%   \vspace{1em}
%   \subfloat[Aggregated throughput vs. number of UEs.]{
%     \includegraphics[width=0.3\textwidth]{figs/agg_rate_ue.png}
%     \label{fig:agg_rate_ue}
%   }
%   \quad
%   \subfloat[Aggregated throughput vs. number of UEs ($P_{O-RU}$).]{
%     \includegraphics[width=0.3\textwidth]{figs/p_ru_agg_rate.png}
%     \label{fig:p_ru_agg_rate}
%   }
%   \quad
%   \subfloat[Aggregated throughput vs. O-RU power.]{
%     \includegraphics[width=0.3\textwidth]{figs/urrllc_embb_p_ru.png}
%     \label{fig:urrllc_embb_p_ru}
%   }
%   \quad
%   \subfloat[Aggregated throughput vs. Power of slice.]{
%     \includegraphics[width=0.3\textwidth]{figs/eu_p_slice.png}
%     \label{fig:eu_p_slice}
%   }
%     \caption{Simulation results showing various performance metrics and algorithm comparisons for SS-VAE, ESA, and DQN models.} %{\color{red} fonts inside figures are too small./Throughput should have dimension.}}
%   \label{fig:simulation_grid}
% \end{figure*}
We evaluated the performance of our SS-VAE method compared to the \ac{ESA} algorithm, focusing on how the number of labeled samples used for training affects the results, as shown in \figref{fig:cmp_alg}. We analyze the aggregated throughput under varying numbers of labeled data points obtained by the SS-VAE algorithm and the \ac{ESA}. As depicted in \figref{fig:cmp_alg}, SS-VAE achieves a similar aggregated throughput compared to the \ac{ESA}. Moreover, the discrepancy between SS-VAE and the \ac{ESA} decreases significantly as the number of labeled data points increases. The difference in aggregated throughput between these methods is $2.1$ for $2000$ labeled samples, which will be decreased using more labeled samples.

We also evaluate the performance of the SS-VAE and \ac{ESA} based on binary variables related to user association ($\alpha_{u_s}^b$) and PRB allocation ($\beta_{u_s,m}^b$). \figref{fig:sum_diff_bin_vars} displays the ratio of the absolute sum of differences in these binary variables between the SS-VAE and \ac{ESA} to the total number of them $(|\alpha_{ESA} - \alpha_{SS-VAE}| + |\beta_{ESA} - \beta_{SS-VAE}|)/\text{\# Number of }(\alpha \& \beta)$, as the number of labeled samples increases. The difference between them reaches its minimum at 2000 samples (4.7\% error) and is expected to decrease significantly as more labeled samples are used.

These results demonstrate the reliable performance of our SS-VAE method in addressing the resource allocation problem. The following section will evaluate its performance under varying initial conditions, including different slice numbers, \acp{UE}, and transmission power and rate thresholds. Although SS-VAE shows lower error than \ac{ESA}, it does not perform identically in resource allocation. Therefore, we will also compare it with the \ac{DQN} algorithm, which has shown effective results in similar resource allocation problems within wireless networks \cite{elsayed2019reinforcement, filali2023communication, joda2022deep}.

\subsection{Performance Results}

\figref{fig:agg_rate_ue} demonstrates the aggregated throughput obtained by SS-VAE, \ac{ESA}, and \ac{DQN} for different number of \acp{UE}. The difference in performance between SS-VAE and \ac{ESA} is consistent across all \acp{UE}. Moreover, ESA achieves higher aggregate throughput than SS-VAE, outperforming \ac{DQN} in total throughput across all \acp{UE}. In addition, the aggregated throughput increases initially as more \acp{UE} join the service for all three examined algorithms. However, it reaches a plateau after 30-35 \acp{UE} per service due to power limitations and interference constraints. \figref{fig:p_ru_agg_rate} shows the aggregated throughput achieved by the SS-VAE algorithm at different maximum power levels of O-\ac{RU} and different numbers of \acp{UE}. In resonation with \cite{motalleb2022resource}, increasing the maximum power of O-\ac{RU} and the number of \acp{UE} leads to higher aggregated throughput. The highest aggregated throughput is observed when the maximum power of O-\ac{RU} is at its highest.

Figures \ref{fig:urrllc_embb_p_ru} and \ref{fig:eu_p_slice} show the aggregated throughput for two service types—\ac{eMBB} and \ac{URLLC}—obtained using the SS-VAE and \ac{DQN} algorithms. Specifically, \figref{fig:urrllc_embb_p_ru} illustrates the weighted aggregate throughput as a function of the maximum power of the \ac{RU}, while \figref{fig:eu_p_slice} presents the throughput variation concerning the maximum power allocated per slice. Both figures compare the performance of SS-VAE and \ac{DQN} across different power constraints for these services. They demonstrate a direct correlation between increased maximum power levels and higher weighted throughput. Across all scenarios, the SS-VAE algorithm consistently outperforms the \ac{DQN} algorithm, providing higher aggregate throughput for both service types, regardless of whether the power levels pertain to the \ac{RU} or the individual slices. Due to their need for higher data rates, \ac{eMBB} service type exhibits higher aggregate throughput than \ac{URLLC} services. This consistent performance advantage underscores the effectiveness of the SS-VAE algorithm in optimizing resource allocation across different power settings.

Although \ac{ESA} ensures a globally optimal solution, its computational complexity makes it impractical for real-world applications. \ac{DQN}, while powerful, faces challenges related to convergence time, exploration efficiency, and sample requirements. Our SS-VAE method employs the strengths of supervised and unsupervised learning to provide a scalable and efficient solution for resource allocation in \ac{O-RAN}, outperforming these benchmarks in real-world scenarios. This contrasts with \ac{DQN}, which may struggle with generalization due to its reliance on extensive exploration and sample efficiency. Additionally, \ac{DQN} often requires significant training time to converge to an optimal solution, especially in environments with high variability and complex constraints. It may also face challenges with exploration efficiency, often requiring many iterations to fully explore the solution space, which can result in suboptimal performance under dynamic conditions.
\section{COMPUTATIONAL COMPLEXITY ANALYSIS} 
\label{complexity}

This section analyzes the computational costs of the \ac{ESA}, \ac{DQN}, and the SS-VAE method and discusses how each algorithm scales with increasing network size and complexity.

\subsubsection{ESA}
The \ac{ESA} evaluates all possible combinations of resource allocations to find the global optimal solution. If there are $|U|$ \acp{UE}, $|B|$ \acp{RU}, $M$ \acp{PRB}, and $P_{\text{levels}}$ number of discrete power levels, the total cost is given by:

$ \text{Total Cost} = (|B|^{|U|}) \times \prod_b (\sum_s \sum_u \alpha_{u_s}^b +1)^M \times P_{\text{levels}}$

Where $\prod_b (\sum_s \sum_u \alpha_{u_s}^b +1)^M$ represents the worst case number of possible \ac{PRB} assignments to \acp{UE} without intra-cell interference in each RU $b$, ESA needs to scale better with increasing network size and complexity due to its exponential growth in computational requirements. Despite its guarantee of finding the global optimal solution, it is impractical for real-time applications and large-scale networks.

\subsubsection{DQN}
The complexity of training a \ac{DQN} is expressed as $O(E \cdot (T \cdot (|S| \cdot |A| + F)))$, where $E$ is the number of episodes, $T$ is the number of steps per episode, $|S|$ is the size of the state space, $|A|$ is the size of the action space, and $F$ denotes the complexity of the neural network forward pass and backpropagation. 
The complexity of making a decision using a trained \ac{DQN} is $O(F)$, where $F$ is the neural network forward pass complexity. 
The total computational cost for training and inference is dominated by the training phase, which can be expressed as:

$\text{Total Cost} = O(E \cdot (T \cdot (|S| \cdot |A| + F)))$

\ac{DQN} can become computationally expensive as the state and action spaces grow due to the need for extensive exploration and training to converge to an optimal policy. This makes \ac{DQN} less scalable for large, complex networks.

\subsubsection{SS-VAE}
The complexity of training a \ac{VAE} is $O(E \cdot (D \cdot L + L^2))$, where $E$ is the number of epochs, $D$ is the dimensionality of the input data, and $L$ is the number of latent dimensions. The complexity of inferring resource allocations using a trained \ac{VAE} is $O(D \cdot L + L^2)$. In addition, the complexity of training with contrastive loss is $O(N \cdot D \cdot L)$, where $N$ is the number of samples. Combining both phases, the total computational cost of the SS-VAE method is:

$\text{Total Cost} = O(E \cdot (D \cdot L + L^2)) + O(N \cdot D \cdot L)$

The SS-VAE method scales efficiently with the number of features and samples due to its reliance on learning from data. While the training phase is computationally intensive, it is performed offline, and the inference phase is relatively lightweight, making it suitable for real-time applications in dynamic O-RAN environments.

\subsubsection{Comparison and Implications}

The \ac{ESA} guarantees finding the global optimal solution, but its computational infeasibility makes it impractical for large-scale networks. \ac{DQN} provides a scalable solution; however, it requires extensive training and help with efficiency in large, complex networks. The SS-VAE method balances computational efficiency and scalability, making it suitable for real-time resource allocation in dynamic \ac{O-RAN} environments. It achieves near-optimal performance with significantly lower computational costs, making it a robust and practical solution for modern \ac{O-RAN} systems.

\section{Conclusions} % Assuming this is a conclusion section
\label{conclusions}

This paper introduced SS-VAE, a novel \ac{GAI} framework for optimizing downlink \ac{O-RAN} resource allocation and network slicing. SS-VAE coordinates \acp{xApp} to manage \ac{RU} associations, power, and \acp{PRB}, maximizing weighted throughput for \ac{eMBB} and \ac{URLLC} services. To tackle the NP-hard optimization challenge, SS-VAE employs a unified semi-supervised \ac{DNN}. This architecture synergistically integrates a \ac{VAE}—whose generative capabilities are honed using \ac{ESA}-derived optimal parameters to determine allocation decisions—with a contrastive loss function that enhances representation learning, generalization, and robustness using both labeled and unlabeled data.

Evaluations against the \ac{ESA} and \ac{DQN} benchmarks demonstrated SS-VAE's ability to achieve near-optimal performance, comparable to \ac{ESA} and significantly outperforming \ac{DQN} while incurring the reduced computational cost. Our simulations confirmed the efficacy of SS-VAE in managing \ac{eMBB}/\ac{URLLC} \ac{QoS} across diverse \ac{O-RAN} scenarios. As a semi-supervised \ac{GAI} framework, SS-VAE offers notable sample efficiency, robust generalization from limited data, and adaptability to dynamic conditions, marking it as a compelling solution for practical \ac{O-RAN} deployments.

Future research will address end-to-end \ac{UE} delay, investigate deployment costs, and incorporate \ac{UE} mobility. We also aim to explore enhanced \ac{GAI} techniques and other advanced \ac{AI} or hybrid optimization algorithms to improve further resource allocation strategies in dynamic \ac{O-RAN} systems, thereby opening new avenues for research in this evolving field.

\bibliographystyle{IEEEtran}
\bibliography{ref.bib}

\end{document}